\documentclass{aa}  

\usepackage{graphicx}
\usepackage{subcaption}
\usepackage{txfonts}
\usepackage{algorithm}
\usepackage{algorithmic}

\newcommand{\bc}[1]{#1}
\newcommand{\bcf}[1]{#1}

\newcommand{\reph}[1]{#1}

\renewcommand*\vec[1]{#1}

\begin{document}

   \title{Deep learning-based deconvolution for interferometric radio transient reconstruction\protect\hphantom{\thanks{Vizualisations of our results are available on the following page: \url{https://github.com/bjmch/DL-RadioTransient}.}}}




\author{Benjamin Naoto Chiche
          \inst{1,2}\and 
          Julien N. Girard
          \inst{3}\and
          Joana Frontera-Pons
          \inst{1}\and
          Arnaud Woiselle
          \inst{2}
          \and
          Jean-Luc Starck\inst{1}
          }

   \institute{Université Paris-Saclay, Université Paris Cité, CEA, CNRS, AIM, Orme des Merisiers, 91191, Gif-sur-Yvette, France\\
              \email{benjamin.chiche@cea.fr}
         \and
             Safran Electronics \& Defense, 100 avenue de Paris, 91344, Massy, France
                     \and
             LESIA, Université Paris Cité, Observatoire de Paris, Université PSL, CNRS, Sorbonne Université, 5 place Jules Janssen, Meudon, 92195, France
             }

   \date{Received 20 September 2022 / Accepted 10 April 2023}

 
\abstract
{
Radio astronomy is currently thriving with new large ground-based radio telescopes coming online in preparation for the upcoming Square Kilometre Array (SKA). Facilities like LOFAR, MeerKAT/SKA, ASKAP/SKA, and the future SKA-LOW bring tremendous sensitivity in time and frequency, improved angular resolution, and also high-rate data streams that need to be processed. They enable advanced studies of radio transients, volatile by nature, that can be detected or missed in the data. These transients are markers of high-energy accelerations of electrons and manifest in a wide range of temporal scales ({e.g.}, from milliseconds for pulsars or fast radio bursts  to several hours or days for accreting systems). Usually studied with dynamic spectroscopy of time series analysis, there is a motivation to search for such sources in large interferometric datasets. This requires efficient and robust signal reconstruction algorithms.} 
  {To correctly account for the temporal dependency of the data, we   improve the classical image deconvolution inverse problem by adding the temporal dependency in the reconstruction problem, and we propose a solution based on deep learning.}
  {We introduce two novel neural network architectures that can do both spatial and temporal modeling of the data and the instrumental response. Then, we simulate representative time-dependent image cubes of point source distributions and realistic telescope pointings of MeerKAT to generate toy models to build the training, validation, and test datasets. Finally, based on the test data, we evaluate the source profile reconstruction performance of the proposed methods and classical image deconvolution algorithm CLEAN applied frame-by-frame.}
  {In the presence of increasing noise level in data frame, the proposed methods display a high level of robustness compared to frame-by-frame imaging with CLEAN. The deconvolved image cubes bring a factor of 3 improvement in fidelity of the recovered temporal profiles and a factor of 2 improvement in background denoising.}
  {The proposed neural networks are not iterative and can benefit from efficient GPU-based architectures. Consequently, they could unlock the development of real-time data processing at the initial cost of learning the behavior of the telescope. Radio observatories are performing sky surveys to produce deep images in ever larger  fields of view, increasing the transient source access window both spatially and temporally. \reph{Our method can effectively capture the temporal structures that are present in such survey data.}}

   \keywords{radio interferometry, transient sources, deep learning, deconvolution}

   \maketitle
%

\section{Introduction}

Next-generation radio facilities like LOFAR~\citep{van2013lofar}, MeerKAT/SKA \citep{booth2012overview}, ASKAP/SKA~\citep{2009IEEEP..97.1482D}, and SKA-LOW \citep{scaife2020big} allow for high spectral, high-rate, improved angular resolutions, and high instantaneous sensitivity. This is a notable improvement for studying transient radio sources via aperture synthesis using radio interferometers. These sources appear and disappear over time and can be randomly distributed in the sky. They are associated with high-energy physical phenomena ({e.g.}, pulsars, rotating radio transients (RRATs), Solar System magnetized objects, and Lorimer-type bursts; \citealt{doi:10.1126/science.1147532}) and, more generally, fast radio bursts (FRBs). Searching for such sources in large datasets produced by these instruments is a new challenge that requires competitive and efficient signal reconstruction algorithms.



Radio interferometers enable imaging via aperture synthesis based on processing the correlations between each pair of antenna signals. In the first approximation, a radio interferometer samples noisy Fourier components of the sky associated with its spatial frequencies (i.e., the visibilities; \citealt{wilson2009tools}) inside the main field of view of the instrument. Under the small-field approximation assumption, the sky can be approximated by computing the inverse Fourier transform of those Fourier samples. 
The number of baselines is limited, so the Fourier map is incomplete. Therefore, it is necessary to solve an ``inpainting''~\citep{garsden2015lofar} inverse problem (i.e., an estimate lacking information in the Fourier plane). Another option is to switch from the inpainting problem in the visibility space to its equivalent deconvolution problem in the image space. Deconvolution of radio images from a static sky in the context of radio interferometry has been the subject of studies for several decades. Notably,   \citet{hogbom1974aperture} designed the original CLEAN algorithm, which is still the most widespread basis for newer deconvolution algorithms in the community. Several variants of this algorithm have been subsequently proposed \citep{schwab1984relaxing,clark1980efficient}. Improvements taking into account source morphology~\citep{Cornwell2008}, spectral dependencies~\citep{rau2011multi}, and sparse representations~\citep{garsden2015lofar,FadiPaper2022,girard2015sparse,dabbech2015moresane,carin2009relationship,carrillo2012sparsity,wiaux2009compressed} have also been investigated in recent years. 
When the sky contains transient sources, classical detection methods rely on frame-by-frame image analysis (e.g., with the LOFAR Transient Pipeline; \citealt{Swinbank_2015}).
However, frame-by-frame transient detection is subject to two observing biases: i) a detection issue when the frames are derived from too short time integration data displaying a high noise level and, conversely, ii) a ``dilution'' problem when time integration is too long to resolve the transient in time, resulting in a time smearing of the transient. Therefore, to account for these biases and sources that possess coherent structure in time,  methods should be designed that directly account for the time-coherent structure (i.e., the source light curve) hidden in the signal. This also occurs in large radio surveys with interferometers. Mapping the visible radio sky requires an optimal use of the observing time and pointing location to reach a target angular resolution and sensitivity. While surveys mainly address the distribution of static sources (astrometry and flux density), transient radio sources might also occur during the short exposure (e.g., $\sim$15 min pointing of MeerKAT) and be missed due to observing biases (detection and dilution). Therefore, finding a robust deconvolution method is key to optimizing both telescope time and transient detectability.

Recently, deep learning (DL) has contributed to achieving far better performance  in solving inverse problems, such as image and video super-resolution, deblurring, and denoising \citep{zhang2018learning,NIPS2014_1c1d4df5,zhang2017beyond,9025527} and computed tomography (CT) and magnetic resonance imaging (MRI) reconstructions \citep{adler2018learned}. In astrophysics, DL has recently been used for tasks such as parametric deconvolution \citep{tuccillo2018deep}, optical image deconvolution \citep{sureau2020deep}, and radio image deconvolution \citep{FadiPaper2022,terris2022image} (and the equivalent inpainting in the Fourier domain; \citealt{schmidt2022deep}). The success of neural networks in these tasks relies on their excellent capability to learn the image prior from data. Based on these premises,  in this study we design novel DL architectures that improve radio image time-series deconvolution in the context of radio transient source reconstruction. The structure of the article is the following. In Section \ref{Sect:Relatedwork} we review studies related to ours. Then, the inverse problem formulation is presented in Section \ref{Sect:Problem}. We introduce new neural network architectures to solve the problem in Section \ref{Sect:Method}. Then, the performance of the proposed networks is evaluated through simulations: Section \ref{Sect:Experiment} details experimental setups and Section \ref{Sect:Results} presents results. Section \ref{Sect:Discussion} discusses hypotheses and limitations. Section \ref{Sect:Conclusion} concludes our work. 

\section{Related work}
\label{Sect:Relatedwork}

\subsection{Deconvolution in radio interferometry}

Two approaches have been classically used in radio interferometry deconvolution: the variational maximum entropy method \citep{frieden1972restoring,narayan1986maximum,wernecke1977two} and the iterative CLEAN algorithm \citep{hogbom1974aperture}. Both methods generally perform well when dealing solely with point sources, but CLEAN is the most widespread technique in the radio astronomy community. This algorithm supposes a finite number of point sources. It restores them based on matching pursuit \citep{bergeaud1995matching} using a single basis vector and the impulse response (hereafter point spread function, PSF) of the telescope that made the observation. \citet{clark1980efficient} proposed a variant of CLEAN by optimizing the algorithm with fast Fourier transform (FFT) and structuring the algorithm computations between major and minor cycles. Minor cycles are carried out in (gridded) image space, whereas major cycles fall in  the ungridded visibility space. Going back and forth between these two spaces led to improvements in both fidelity and accuracy. This strategy was further developed in \citet{schwab1984relaxing}.  \citet{Cornwell2008} and \citet{rau2011multi} brought in further improvements by respectively taking into account the morphological and spectral behavior of the sources. More recently, several teams have addressed the deconvolution problem within the compressed sensing framework \citep{garsden2015lofar,girard2015sparse,dabbech2015moresane,carin2009relationship,carrillo2012sparsity,wiaux2009compressed,mingjiang}, and then DL~\citep{terris2022image}.  \citet{schmidt2022deep} also used DL, but solved the equivalent inpainting problem in the Fourier space. \bcf{However, most of these methods do not take into account the temporal structures of time-evolving sources in the context of image time-series deconvolution. 
The method presented in \citet{mingjiang} is the only existing method that considered a temporal prior. However, this method is extremely slow in practice, and struggles when dealing with interferometric images with realistic dimensions. Moreover, it introduces many false detections in empty regions of the reconstructed skies. For these reasons, we do not consider it in our empirical study that is presented later in this paper.}




\subsection{Deep learning  and inverse problems}

DL has shown remarkable results in solving inverse problems in image and video restoration. In inverse problems, the aim is to reconstruct the original data $x$ from some observed data $y$. They are related through a degradation model with the form $y = D_\alpha (Ax)$~\citep{pustelnik:hal-01164833}, where   matrix $A$ models linear transformations (such as blurring) and   $D_\alpha$ models other degradations. This function is parameterized by $\alpha$. First-generation networks solving inverse problems \citep{NIPS2014_1c1d4df5,dong2015image,kim2016accurate} approximate $x$ by learning in a supervised manner the inverse function $\hat{x} = N(y; \theta)$ from the data $y$, where $\theta$ are parameters of the network. These approaches perform well when the degradation model has fixed parameters (i.e., $A$ and $\alpha$ are unique across the train and test data). Their performance deteriorates in a multiple degradation scenario (i.e., when $A$ and $\alpha$ vary across the train and test data). In the case of radio interferometric observation, the blurring operator $A$ depends on the time of each image and on the considered location in the sky, and thus changes between samples. \citet{zhang2018learning} improved upon the first-generation networks by proposing a single convolutional neural network (CNN) that can handle multiple degradations. This network learns the mapping $\hat{x} = N(y, A, \alpha; \theta)$ from the data. More recently,  \citet{gu2019blind} proposed another way to handle multiple degradations based on the spatial feature transform (SFT) layer introduced in \citet{wang2018recovering}. This layer provides an affine transformation of its input feature maps conditioned on the knowledge about $A$ and $\alpha$. 





Other strategies that can handle multiple degradations use iterative approaches, often derived from convex optimization algorithms, coupled with neural networks. We broadly distinguish two methods. One of them is Deep Unrolling, which consists in unfolding the iterative loop of a classical iterative algorithm with a given number of iterations and representing all operations as layers of a neural network~\citep{zhang2020deep,adler2018learned,9363511,diamond2017unrolled,chiche2020deep}. The network can be trained and optimized based on data, while keeping the knowledge of the degradation in its internal structure. The network is notably factorized into a prior step that encodes learned image prior and a data step which leverages knowledge about the degradations. The other one is \bc{Plug \& Play, which uses an iterative proximal algorithm, for instance in the framework of alternating direction method of multiplier plug \& play~\citep{venkatakrishnan2013plug,7542195,7744574} or regularization by denoising~\citep{romano2017little,reehorst2018regularization}, and replaces the operator related to the prior by a denoising CNN~\citep{meinhardt2017learning,NIPS2017_38913e1d,gupta2018cnn,terris2022image} or a series of denoising CNNs trained in different settings~\citep{zhang2017learning}.
    }


However, the former can encounter a memory issue at training time if we want to unroll an important number of iterations. Moreover, its data step generally applies the linear operator $A$ and its adjoint operator $A^T$ to the input signal, diluting the signal too much if the operator is strongly ill-conditioned. The latter approach shares a drawback with classical iterative algorithms like CLEAN: it is slow because it requires an important number of iterations to deconvolve an image.

\section{Problem}
\label{Sect:Problem}

\subsection{Interferometry imaging problem} \label{imaging problem}


This study deals with imaging by aperture synthesis from interferometric data. The limited number of antennas and observing baselines, time, and frequencies restrict the amount of accessible samples of the sky visibility function. In addition, these visibilities are subject to  noise that can be modeled as an additive Gaussian noise in the first approximation. In a limited field of view and ignoring direction-independent or direction-dependent effects and calibration issues, this ill-posed inverse problem can be expressed in the Fourier space (i.e., the measurement space) as 

\begin{equation}
    V_y = M(V + \epsilon)
    \label{fourier}
,\end{equation}

\noindent where $V_y$ is the collection of observed visibilities, $V$ is the true visibility function, $\epsilon$ is approximated as an additive white Gaussian noise, and $M$ is a sampling mask representing the limited access of an interferometer to the measurement (depending on   antenna  configuration and observational parameters).

Equation~(\ref{fourier}) can be rewritten as a deconvolution problem formulated in the image or direct space. This problem links the observed degraded sky image to the corresponding true sky image

\begin{equation}
    y = h * x + \eta
    \label{direct}
,\end{equation}





\noindent with $\mathcal{F}^{-1}(V_y) = y$, $\mathcal{F}^{-1}(M) = h$, $\mathcal{F}^{-1}(V) = x$, and $\mathcal{F}^{-1}(M) * \mathcal{F}^{-1}(\epsilon) = \eta$, and where $*$ denotes the convolution operation; $y$ is the observed image, called the {dirty image}; $h$ is the PSF, also called the {dirty beam}, which represents the sampling operation in Fourier space by the interferometer; and $x$ is the ground truth (GT) image. We note that  $\sigma_{\epsilon} $ is the noise level of $\epsilon$. Thus, the corresponding variance for $\eta$ can be obtained from $\sigma = ||M||_2 \cdot \sigma_\epsilon = ||\mathcal{F}(h)||_2 \cdot \sigma_\epsilon$.


\subsection{Extension to transient imaging}
\label{video problem}

To enable robust imaging of transient sources, instead of solving frame by frame, which can be subject to observation bias, we extend the problem in Eq.~(\ref{direct}) to a deconvolution problem accounting for the temporal dependency of the different terms (i.e., sky, noise, and instrument sampling):

\begin{equation}
    y_t = h_t * x_t + \eta_t, \qquad  
    t \in I = \{t_0, \ldots, t_{T-1}\}  
    \label{transient imaging}
.\end{equation}

\noindent Here $\mathcal{F}^{-1}(V_{y, t}) = y_t$, $\mathcal{F}^{-1}(M_t) = h_t$, $\mathcal{F}^{-1}(V_t) = x_t$, and $\mathcal{F}^{-1}(M_t) * \mathcal{F}^{-1}(\epsilon_t) = \eta_t$. The noise level of $\eta_t$ is $\sigma_t = ||M_t||_2 \cdot \sigma_\epsilon = ||\mathcal{F}(h_t)||_2 \cdot \sigma_\epsilon$. By stacking $\{y_t\}_{t \in I}$, $\{x_t\}_{t \in I}$ and $\{h_t\}_{t \in I}$ in the temporal dimension, we respectively obtain a {dirty cube}, a GT cube and a PSF cube. These cubes are three-dimensional data structures denoted $Y$, $X$, and $H$, respectively; $I$ is the set of $T$ time steps ordered following the observation intervals. 
In the single image problem~(\ref{direct}), $h$ depends on the total time and frequency integration of observation, while the sky $x$ is supposed to be static. As the sky naturally rotates over the instrument during the observation, the morphology of $h$ depends on the interferometer location on Earth, declination of the source, and observing dates. In the dynamic imaging problem extension~(\ref{transient imaging}), the $h_t$ operator has a time dependency (i.e., the instrumental response varies for consecutive single observing time intervals). As a result, both the sky and the interferometer responses vary over time. The associated mask $M_t$ samples the Fourier transform of the sky at different time dates, enabling the possibility of capturing the temporal evolution of the observed sky.



We assumed that the datacube $X$ only contains point-like sources for simplification. We assume each source has an angular scale that is  much smaller than the angular resolution brought by the PSF. In addition, we assume that the cube contains mixtures of sources with constant and varying flux densities over time. Their locations on the sky will be random but constant during the observation.

Because a radio interferometric dataset provides exact information on the baseline length and orientation for all samples, the morphology and time dependency of $h_t$ are known for all $t$.
Finally, we control the noise level $\sigma_{\epsilon}$ to mimic the various quality levels of the observations. Following the discussion in Section \ref{Sect:Relatedwork},  we are thus in a  multiple degradation scenario, as in~\citet{zhang2018learning}.



In this study, for the sake of simplification, we focus on the monochromatic case where the observing frequency is fixed. We therefore do not deal with the dependency in frequency of the imaging problem.

\section{Method}
\label{Sect:Method}



In the context of the time-series image deconvolution problem ~(\ref{transient imaging}),  a single image deconvolution method, such as CLEAN,   can be independently applied in a frame-by-frame approach. However, this method does not capture the temporal structure of the sky. A method capable of dealing with the temporal evolution of both telescope and the sky is required. Given recent successes of neural networks in various restoration tasks, we propose to solve  problem~(\ref{transient imaging}) based on DL. In our setting, a network realizes the following mapping: $\hat{X} = N(Y, H, \sigma_\epsilon ; \theta) $. 
Its input contains a degraded cube and information about the degradation $(H, \sigma_\epsilon)$ in a non-blind manner. The symbol $\theta$ denotes parameters of the network that are learned from training data $\{Y_i, X_i, H_i, \sigma_{\epsilon, i} \}$. We propose two implementations of the network $N$, called 2D-1D Net and Deflation Net, respectively.


\subsection{Multiple degradations}

We adopt the following scheme to incorporate knowledge about the image formation model and handle multiple degradations. Each $h_t$ in $H$ is originally of size $l \times l $ (with $ l=256 $ in our study; see Section~\ref{Sect:Experiment}), but is first center-cropped with crop size $ r \times r $ ($ r = 96 $ in our study) and projected onto a $b$-dimensional linear space by a PCA projection matrix $ P \in \mathbb{R}^{b \times r ^2} $. The matrix $P$ is learned from all PSFs that constitute the training PSF cubes. \bc{In our empirical study (which is detailed in Sects.~\ref{Sect:Experiment} and~\ref{Sect:Results}), the value $b = 50$ explains 90 \% of the total variance. The higher and the closer to 1 this percentage is, the better it is (for example,  \citealt{zhang2018learning} preserved about 99.8\% of the energy of their training blurring kernels with PCA), but a value of $b$ higher than 50 makes our DL models (detailed in the following) too large in terms of number of parameters and provokes a memory error at training time}. We denote as $\vec{h}_t ^ b$ this projected PSF, and  $\vec{h}_t ^ b$ is then concatenated with $\sigma_t = ||M_t||_2 \cdot \sigma_\epsilon = ||\mathcal{F}(h_t)||_2 \cdot \sigma_\epsilon $. We denote this vector $\vec{h}_t^{b+1} $.


\subsection{2D-1D Net}

\begin{figure}[!htbt]
\centering
\includegraphics[width=0.5\textwidth]{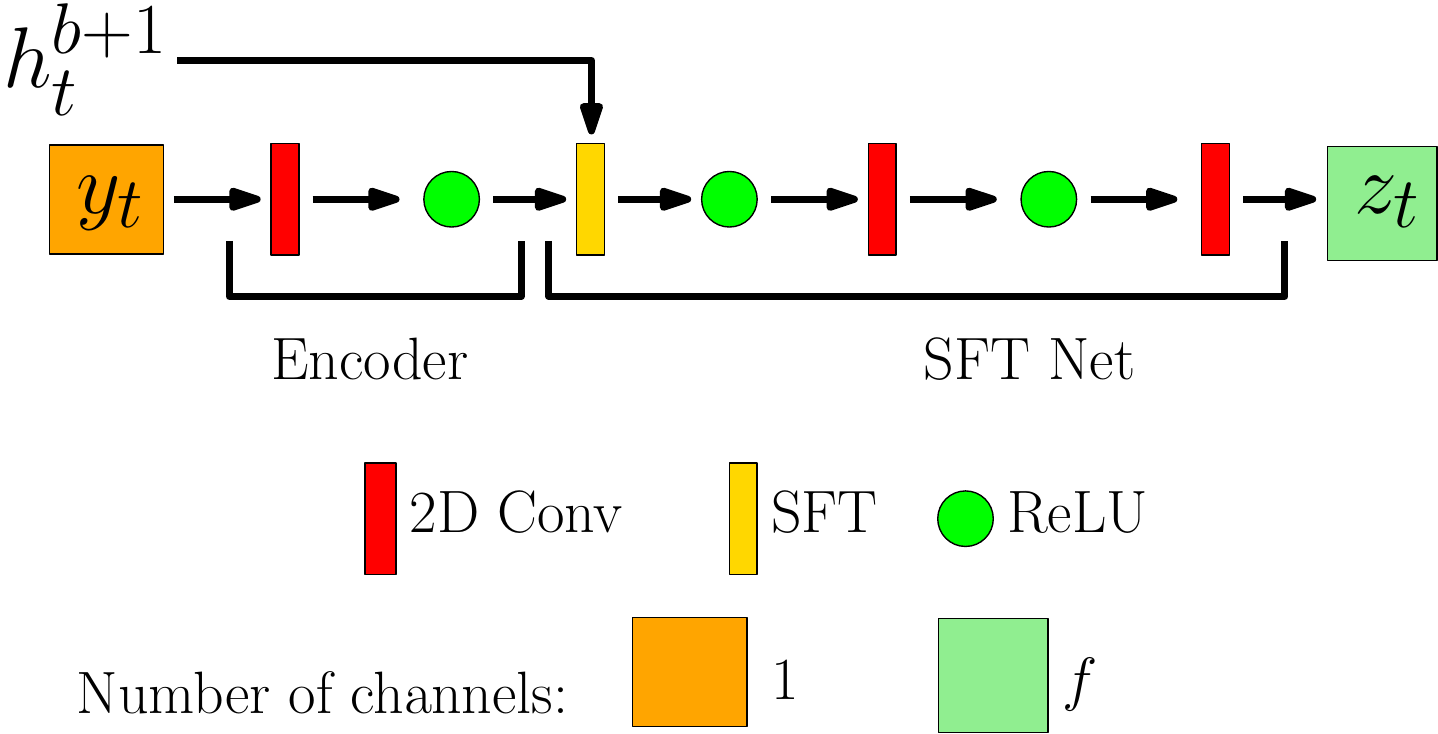}
\caption{2D Net. Each convolutional layer outputs $f$ feature maps. $f = 32$ in our study. The kernel size of each convolutional layer is set to $3 \times 3$.} 
\label{2D}
\end{figure}

\begin{figure}[!htbt]
\centering
\includegraphics[width=0.5\textwidth]{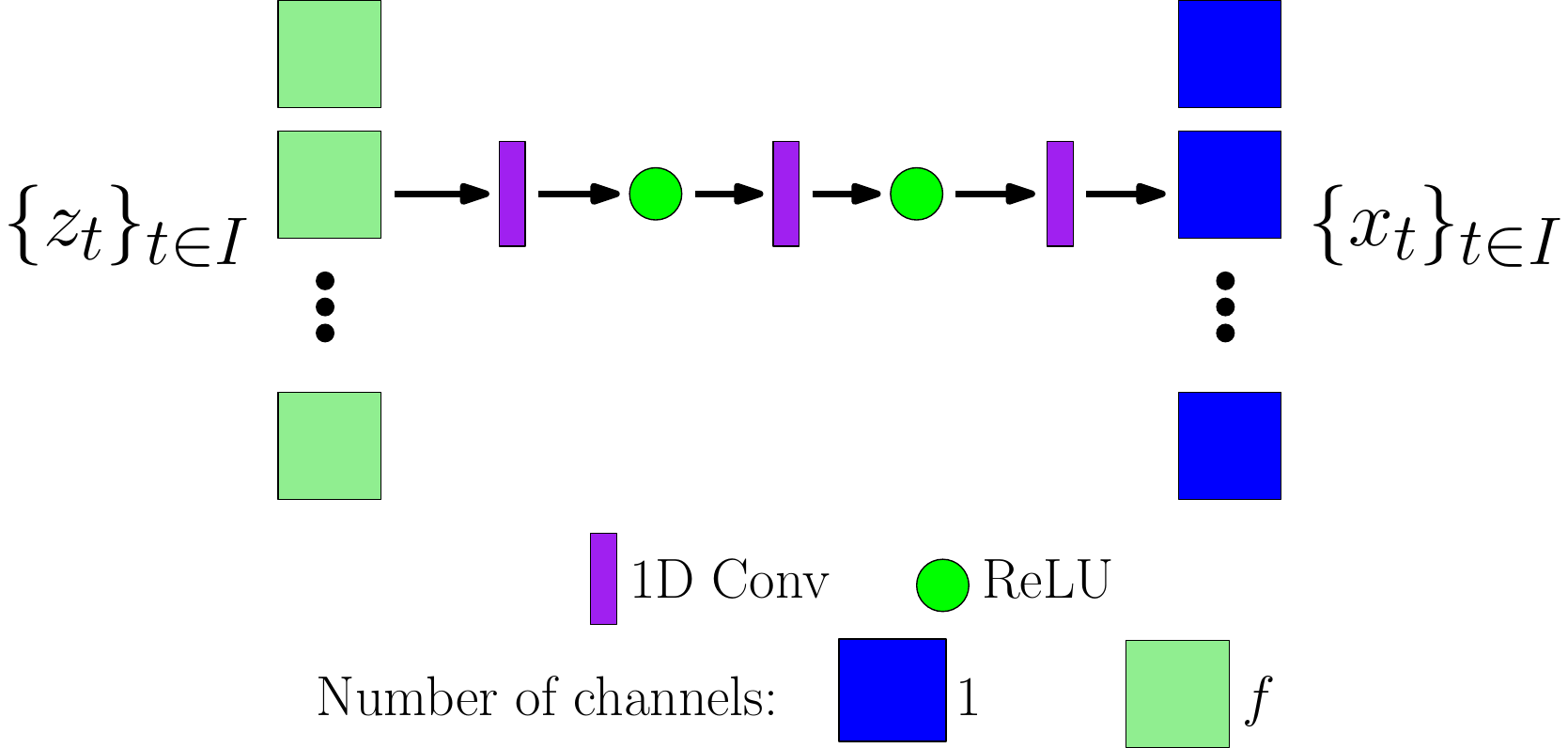}
\caption{1D Net. $\{z_t\}_{t \in I}$ has dimensions $ \bc{f} \times T \times \bc{height} \times \bc{width} $. Each convolutional layer outputs $f$ feature maps, except the last,  which outputs images with one  channel each. Each 1D convolutional layer has   kernel size $5$.}
\label{1D}
\end{figure}

\begin{algorithm}[!htbt]
\caption{2D-1D Net. $C_r$ denotes the operation that center-crops a 2D structure with crop size $r$.}
\label{alg:algorithm2D1D}
\textbf{Input}: $(Y, H, \sigma_\epsilon) = (\{y_t\}_{t \in I}, \{h_t\}_{t \in I}, \sigma_\epsilon) $\\
\textbf{Output}: $\hat{X}$
\begin{algorithmic}[1] 
\STATE  $ S \leftarrow \textrm{an empty list}$
\FORALL{$t \in [|0, T-1|]$} \label{loop}
\STATE  $ \vec{h}_t ^ b \leftarrow P C_r h_t $
\STATE $\sigma_t = ||M_t||_2 \cdot \sigma_\epsilon = ||\mathcal{F}(h_t)||_2 \cdot \sigma_\epsilon  $  
\STATE  $ \vec{h}_t^{b+1} \leftarrow \textrm{Concat}(\vec{h}_t ^ b, \sigma_t) $
\STATE $f_t \leftarrow \textrm{Encoder}(y_t)$
\STATE $z_t \leftarrow \textrm{SFT Net} (f_t, \vec{h}_t^{b+1})$ \label{z}
\STATE $S.\textrm{append}(z_t)$ \label{zlist}

\ENDFOR
\STATE $Z \leftarrow \textrm{StackAlongTimeAxis}(S)$ \label{stack}
\STATE $\hat{X} \leftarrow \textrm{1D Net}(Z)   $ \label{recons}
\STATE \textbf{return} $\hat{X}$
\end{algorithmic}
\end{algorithm}



We decouple $N$ into two modules that sequentially process the input data. The first module is a network with 2D convolutional layers that successively and independently encode each image in the degraded cube into feature maps. Each of these encodings considers the PSF and the noise level used in the degradation. This transformation performs a deconvolution of the degraded image. This module is referred to as 2D Net. After these independent deconvolutions, the produced feature maps are stacked in an additional temporal dimension and given to the second module. This module captures temporal structures within the stacked maps and estimates the GT cube. The temporal profile of a source is continuous in time, which justifies this architectural choice. Because point sources do not spatially move in our study, the temporal structure is extracted based on 1D convolutional layers along the time dimension. We call this module 1D Net. By unifying the two modules, we build the entire network 2D-1D Net.

Algorithm~\ref{alg:algorithm2D1D} summarizes the 2D-1D Net. In the first place, each pair $(y_t, \vec{h}_t^{b+1})$ is given to 2D Net to produce an intermediate feature map $z_t$ (lines~\ref{loop} to~\ref{zlist}). Figure~\ref{2D} describes the structure of this subnetwork. 
This network is composed of an encoder that extracts input features and SFT Net that can manage multiple degradations. In this step, 2D Net deconvolves $y_t$ by using $\vec{h}_t^{b+1}$. The SFT layer uses this vector to modulate feature maps. This layer was introduced in~\citet{wang2018recovering} and used in~\citet{gu2019blind} for the first time to handle multiple degradations in inverse problems. This layer applies an affine transformation to the feature maps conditioned on the degradation maps $F_t^{(h)}$, which are obtained by stretching $\vec{h}_t^{b+1} $ into size $(b+1) \times \bc{height} \times \bc{width} $, where all the elements of the $i$-th map equal the $i$-th element of $\vec{h}_t^{b+1}$. \bc{Here, $ height $ and $ width$ respectively indicate the height and width of input images and feature maps. We   see that in our experiments, $ height = width = 256$}. The affine transformation involves scaling and shifting operations

\begin{equation}
\textrm{SFT}(F_{in},F_t ^{(h)}) = \gamma \odot F_{in} + \beta    
,\end{equation}

\noindent where $\gamma$ and $\beta$ are estimated by additional small convolutional layers and $\odot$ is the Hadamard product. 

Next, $\{z_t\}_{t \in I}$ are stacked in the temporal dimension. This gives Z, a tensor of dimensions $ \bc{f} \times T \times \bc{height} \times \bc{width}$ (line~\ref{stack}). The output is passed to 1D Net, composed of three 1D convolutional layers working along the temporal dimension, interlaced with ReLU activations (line~\ref{recons}). The kernel size of each 1D convolutional layer is set to $5$. With three such layers, the temporal receptive field is $5 + 4 + 4 = 13$, which is enough considering the temporal extension of the experimental transient events in this work (see Section~\ref{transient profiles}). Figure~\ref{1D} details the 1D Net architecture.

\subsection{Deflation Net}





Algorithm~\ref{alg:algorithmDeflation} summarizes the Deflation Net. This network is based on the same subnetworks as 2D-1D Net but involves a different computation flow. Specifically, it decouples the reconstruction of constant and transient sources. In the first place, both input images $\{y_t\}_{t \in I}$ and PSFs $\{h_t\}_{t \in I}$ are averaged over the temporal dimension (lines~\ref{average image} and~\ref{average psf}). The average input image presents a reduced noise level (computed in line~\ref{average noise}) compared to each image in the input cube. The average PSF is better conditioned than each PSF of the PSF cube. Then, the average image is deconvolved in the feature space based on the average PSF to give the average sky features (line~\ref{average sft}). Next, this average sky is reconvolved by the corresponding PSF (line~\ref{subst}) and subtracted to the individual degraded sky in the feature space at each time step. Each resulting image only contains transient sources. This sky is then deconvolved based on the individual PSF to reconstruct transient sources (line~\ref{indconv}). This individual deconvolved image and the average deconvolved sky are finally summed in the feature level via skip connections (line~\ref{summ}). This processing is done for each time step, and all outputs are then sent to the final 1D Net (lines~\ref{sta} and~\ref{send}).


\begin{algorithm}[!htbt]
\caption{Deflation Net. $C_r$ denotes the operation that center-crops a 2D structure with crop size $r$.}
\label{alg:algorithmDeflation}
\textbf{Input}: $(Y, H, \sigma_\epsilon) = (\{y_t\}_{t \in I}, \{h_t\}_{t \in I}, \sigma_\epsilon) $\\
\textbf{Output}: $\hat{X}$
\begin{algorithmic}[1] 
\STATE $\underline{y} \leftarrow \frac{1}{T} \sum_{t=0}^{T-1} y_t$ \label{average image}
\STATE $\underline{h} \leftarrow \frac{1}{T} \sum_{t=0}^{T-1} C_r h_t$ \label{average psf}
\STATE $\underline{\vec{h}} ^ b \leftarrow P \underline{h} $ 
\STATE $\underline{\sigma} \leftarrow \frac{1}{T} (\sum_{t=0}^{T-1} ||M_t||_2 ^2 \sigma_\epsilon ^ 2)^{1/2} = \frac{1}{T} (\sum_{t=0}^{T-1} ||\mathcal{F}(h_t)||_2 ^2 \sigma_\epsilon ^ 2)^{1/2} $ \label{average noise}
\STATE $\underline{\vec{h}}^{b+1} \leftarrow \textrm{Concat}(\underline{\vec{h}} ^ b, \underline{\sigma})$
\STATE $\underline{f} \leftarrow \textrm{Encoder} (\underline{y})$ \label{average feat}
\STATE $\underline{z} \leftarrow \textrm{SFT Net} (\underline{f}, \underline{\vec{h}}^{b+1})$ \label{average sft}
\STATE  $ S \leftarrow \textrm{an empty list}$
\FORALL{$t \in [|0, T-1|]$}
\STATE  $ \vec{h}_t ^ b \leftarrow P C_r h_t $
\STATE $\sigma_t = ||M_t||_2 \cdot \sigma_\epsilon = ||\mathcal{F}(h_t)||_2 \cdot \sigma_\epsilon  $
\STATE  $ \vec{h}_t^{b+1} \leftarrow \textrm{Concat}(\vec{h}_t ^ b, \sigma_t) $
\STATE $f_t \leftarrow \textrm{Encoder}(y_t)$
\STATE $\Delta_t \leftarrow f_t - h_t * \underline{z}$ \label{subst}
\STATE $d_t \leftarrow \textrm{SFT Net} (\Delta_t, \vec{h}_t^{b+1})$ \label{indconv}
\STATE $z_t \leftarrow d_t + \underline{z}$ \label{summ}
\STATE $S.\textrm{append}(z_t)$

\ENDFOR
\STATE $Z \leftarrow \textrm{StackAlongTimeAxis}(S)$ \label{sta}
\STATE $\hat{X} \leftarrow \textrm{1D Net}(Z)   $ \label{send}
\STATE \textbf{return} $\hat{X}$
\end{algorithmic}
\end{algorithm}

\section{Experiment}
\label{Sect:Experiment}

\begin{figure*}[!h]
\centering
\includegraphics[width=2\columnwidth]{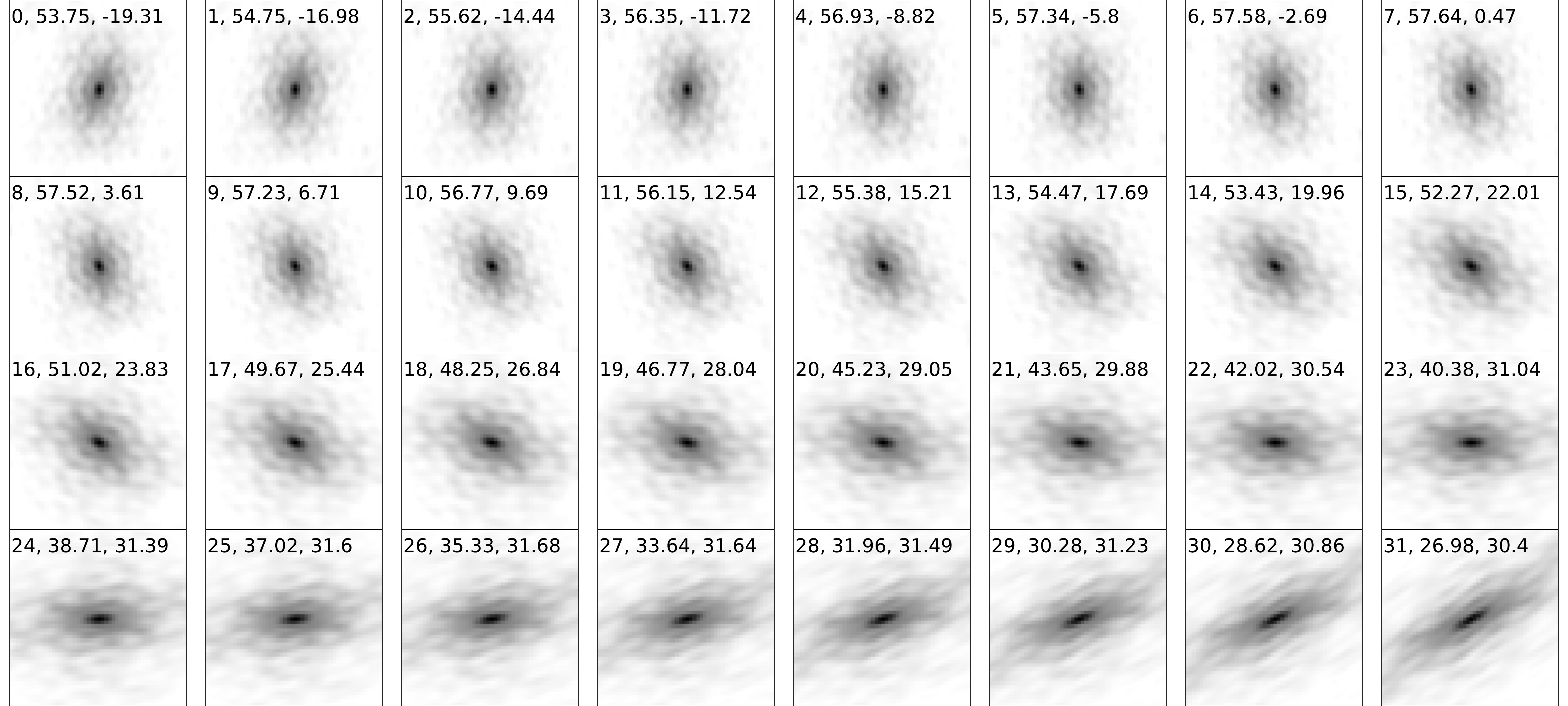}
\caption{The PSF test cube chosen for evaluation. The text in each panel reports: time step, elevation in degrees, azimuth in degrees. Each PSF is normalized to 1 (black). The gray color scale was reversed for clarity. \bcf{The apparent rotation of the PSF with time is the result of the projection of the interferometer aperture toward the source direction, which depends on the interferometer location, declination of the source, and observing time (see Figure~\ref{fig:coordinates} for an illustration).}
}
\label{testpsf510}
\end{figure*}

\subsection{Datasets}

We generate disjoint training, validation, and test sets of GT and PSF cubes  at various noise levels. The corresponding dirty cubes are generated based on Eq.~(\ref{transient imaging}).

\subsubsection{PSF cubes}

\begin{figure}[!h]
\centering
\includegraphics[width=1\columnwidth]{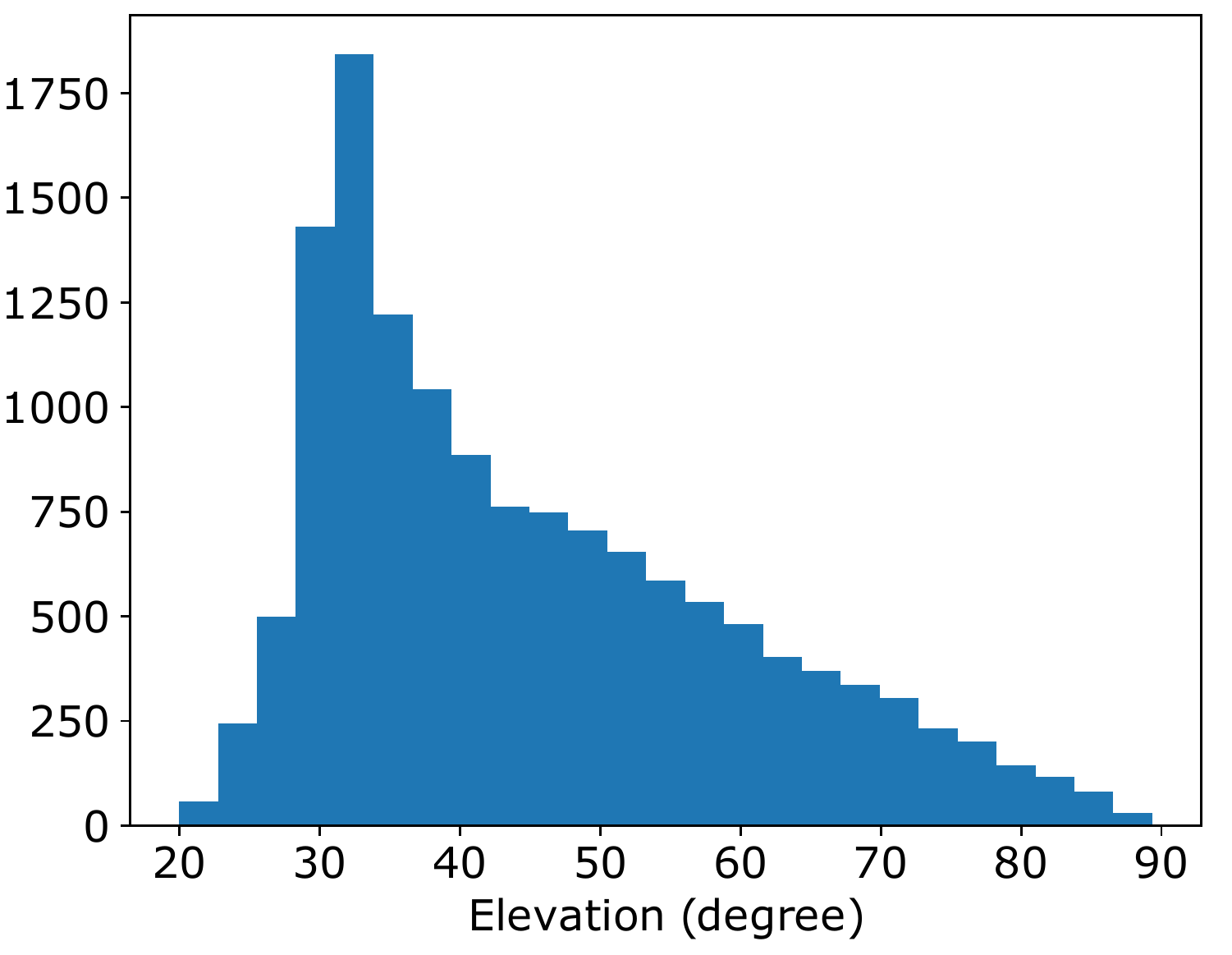}
\caption{Histogram of distribution of elevations in the training PSF set. The distribution is skewed toward the south celestial pole where circumpolar sources can be observed from MeerKAT.}
\label{hist}
\end{figure}






We simulate the interferometric response of MeerKAT in the L-band, using its current 64-antenna distribution. The observing frequency is fixed to 1420 MHz. The location of MeerKAT on Earth is $Long=21.33^\circ$, $Lat=-30.83^\circ$, $h=1195 m$. The minimum pointing elevation is set to $20$ degrees. Based on these parameters, we randomly pick PSF cubes by sampling the local visible sky for a given observing time and observing rate. Each cube accounts for the typical eight-hour tracking observation of a randomly selected source in the J2000 reference frame that is visible in the local sky during the eight hours. This period is divided into successive intervals of 15-minute scans. They produce a 15-minute integrated PSF associated with the (u,v) coverage computed toward the source's current location. Each PSF is projected into a spatial grid of $256 \times 256 $ pixels. Hence, the cube has the dimensions $32 \times 256 \times 256 $. Adopting the notations from Section~\ref{video problem}, we have $T=32$ and $t_{i+1} - t_{i} = 15 $ minutes. We generated 435 training, 50 validation, and 50 test PSF cubes samples. For all of them the source elevations at the beginning of the observation are above 20 degrees as seen from MeerKAT. Figure \ref{testpsf510} depicts the 32 (15 min) frames of a  PSF cube. As the source appears to be rotating above the interferometer during the eight-hour observation, the projected baselines vary, leading to a continuous rotation and warping of the PSF.




Figure \ref{hist} depicts the effective distribution of elevation in training PSFs set. This distribution is skewed toward 30$^\circ$ elevation in the case of the telescope at hand, MeerKAT. This is expected because the PSF simulation accounts for the visibility of sources, in the southern sky, as seen from MeerKAT. For any given observation duration (e.g., 8h max) and integration (e.g., 15 min / image), we computed the telescope source tracking and its aperture projection, both required to derive the effective (u,v) coverage and therefore the PSF. An astrophysical source is associated with  a unique pair of coordinates (declination $\delta$ and right ascension $\alpha$) on the celestial sphere (see Figure~\ref{fig:coordinates} for an illustration). Depending on the source declination ($\delta$, in the equatorial frame), hour angle (associated with right ascension $\alpha$ and time of observation), and observation duration, not all ($\alpha$,$\delta$) directions are accessible. Therefore, a uniform random distribution of directions, drawn from the accessible direction window in the sky, will appear skewed around the local elevation of the south celestial pole (SCP), located around 30$^\circ$ for MeerKAT. Circumpolar sources close to the SCP are always accessible.

\begin{figure}[!htbt]
\centering
\includegraphics[width=\columnwidth]{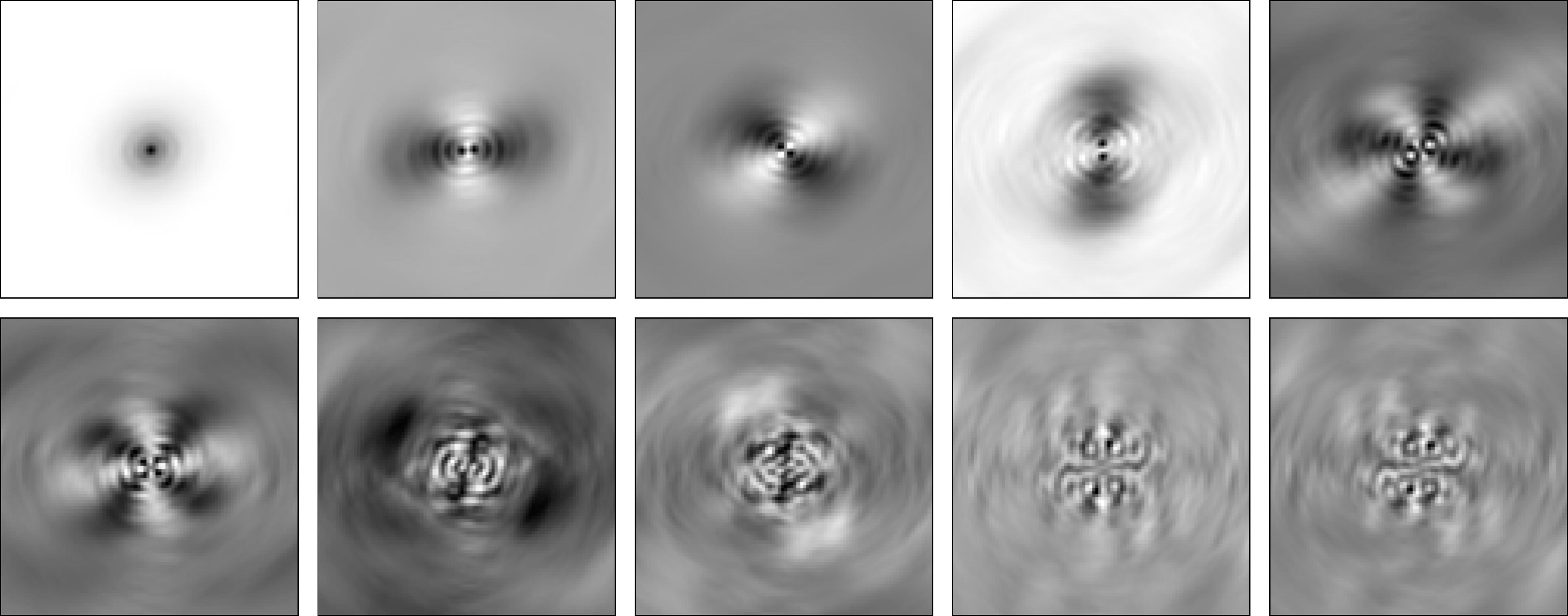}
\caption{PCA eigenvectors for the first ten largest eigenvalues computed over the whole PSF set.}
\label{pca}
\end{figure}

The PCA projection matrix $P$ that encodes knowledge about the PSF 
is learned from all PSFs of the train PSF set (i.e., from the $435 \times 32 =  13920$ PSFs). The value of PCA components $b = 50$ explains 90 \% of the total variance. Figure~\ref{pca} shows the ten main PCA eigenvectors that composed the PSF of the whole set.


\subsubsection{GT cubes} \label{transient profiles}


We suppose the sources to be unresolved points sources in the sky image. The size of the pixel on the sky is fixed to $1.5''$. We assume that within a sky image of size $ 30' \times 30' $ (i.e., $ 1200 \times 1200 $ pixels), at most 30 constant and 2 transient point sources can be placed. This distribution of sources can be considered compatible with shallow imaging like that of  MeerKAT. 
Following this distribution, we generate 39000 training sky cubes with dimensions $ 32 \times 256 \times 256 $. These data are divided into three equal parts, containing zero, one, or two transient sources per field. Each validation and test set contains 66 cubes following the same distribution of sources, with half of them containing a transient source and the other half containing two. The source peak flux density (i.e., the value of the pixel of a constant source in the ground truth) is randomly sampled between 1 and 100. Regarding a transient source, its amplitude is a discrete function $  A(t_i)  $ sampled from the continuous functions $ A(t) $. For each source, this profile is randomly chosen between the following models: {gate}, {gaussian}, and {skew-normal}. Their parameters are randomly chosen, and the transient source's maximum amplitude is randomly chosen between $[50, 100]$. Examples of temporal profiles are shown in Section~\ref{Sect:Results}, Figure ~\ref{1transit}. \bcf{Our neural network systems consider normalized pixel values (i.e., the GT cube’s pixel values are divided by 100) so that they lie between 0 and 1.}

\subsection{Training procedure}

From each GT cube in the training, validation, and test sets, the corresponding \emph{dirty cube} is generated based on Eq.~(\ref{transient imaging}). Mini-batch gradient descent with a batch size of 4 is used for training. For each example in the batch, a train PSF cube is randomly picked, and the noise level $\sigma_\epsilon$ is randomly sampled from 
$[0, 6]$. Data augmentation with random flipping and/or transposition was performed. The learning rate is set to $10^{-4}$, and we train our models for 100 epochs each. The loss function is the pixel-wise mean-squared error (MSE) between GT and estimated cubes.  \citet{vafaei2019deepsource} claimed that if GT images are skies with point sources, then the MSE loss function would be suboptimal and instead proposed to smooth the point sources. However, our study obtained satisfactory empirical results with this pixel-wise MSE between GT and estimated cubes.



\subsection{Evaluation metrics}

Given a constant or transient source at a certain location in a GT cube $X$, we can first define a subcube obtained by locally cropping the cube around the source with a patch size $p \times p $ ($p=3$ in our study, defining the region $\mathcal{D}_s$). We can also extract a subcube at the same location for the estimated cube. We can then compute the root MSE (RMSE) between the two subcubes. This error quantifies the fidelity of the restored temporal profile. The GT subcube norm can normalize this value. We denote as $\mathrm{NRMSE_s}$ this error for the source $s$.

To measure the input signal quality, for each source $s$ in the {dirty cube} $\{y_t\}_{t \in I}$ with location $(i_s, j_s)$ we evaluate its signal-to-noise ratio $\mathrm{S/N_s}$. We adopt the following definition:

\begin{equation}
\centering
\mathrm{S/N_s} = \displaystyle \sum_{i,j \in \mathcal{D}_s}\frac{y_\tau [i, j]}{\sigma_b} 
.\end{equation}

\noindent Here $\tau$ refers to the temporal localization of the transient. It corresponds to the time step when the amplitude of the transient source is maximum in the GT cube. For a constant source, we set $\tau = 15$. $\mathcal{D}_s$ is the local patch of size $3 \times 3$ centered on the true source location $(i_s, j_s)$. The parameter $\sigma_b$ denotes the background noise estimated on $y_\tau$ after excluding $\mathcal{D}_s$. In radio images, Peak S/N is the ratio of peak flux density (usually a single pixel) of the source to the local noise root mean square. Here, to absorb pixel gridding bias, the source flux is accounted over the region $\mathcal{D}_s$. This   ensures that all the recovered flux density of the source is properly accounted for in the  presence of noise. This error could lead to detrimental and unfair representations of light curve reconstruction with different methods.

Moreover, to measure the performance of background denoising (i.e., restoration of the empty region of the sky) we exclude all $\mathcal{D}_s$ subcubes for a GT cube. We operate the same procedure on the corresponding estimated cube and compute the RMSE between the two. We note this metric $\mathrm{RMSE}_\mathrm{noise}$.

We compare the proposed algorithms for the cube restoration against frame-by-frame CLEAN deconvolution of the {dirty cubes}. To avoid biasing the final result during the CLEAN final restoring beam step, we considered only the detected CLEAN components associated with the detected sources. PSF cube and noise level are provided to CLEAN in a non-blind manner. They intervene in defining a threshold from which the algorithm iteration stops. Furthermore, to prevent bias on the use of CLEAN due to a high-noise environment of the {dirty cube}, we stopped CLEAN based on a maximum number of iterations and, as for other reconstruction methods, also considered only counting the flux density around the source location in $\mathcal{D}_s$.



\section{Results}
\label{Sect:Results}



\subsection{Fixed test PSF cube and varying noise} \label{fixedpsf}


We picked a PSF test cube and evaluated the methods on the test sky cubes with the following noise levels: $\sigma_\epsilon \in  \{0, 1, 2, 3, 4, 5, 6\} $. The injected noise levels were selected to range from low noise level cases, where the constant and transient sources are readily detectable in the \emph{dirty cubes}, to high noise level cases, where no sources can be seen (such a high noise level is displayed on the first line of Figure \ref{18}). Figure~\ref{testpsf510} shows the PSF test cube that has been picked. We see that the PSF rotates with time.



\begin{figure}[!h]
\centering
\includegraphics[width=1\columnwidth]{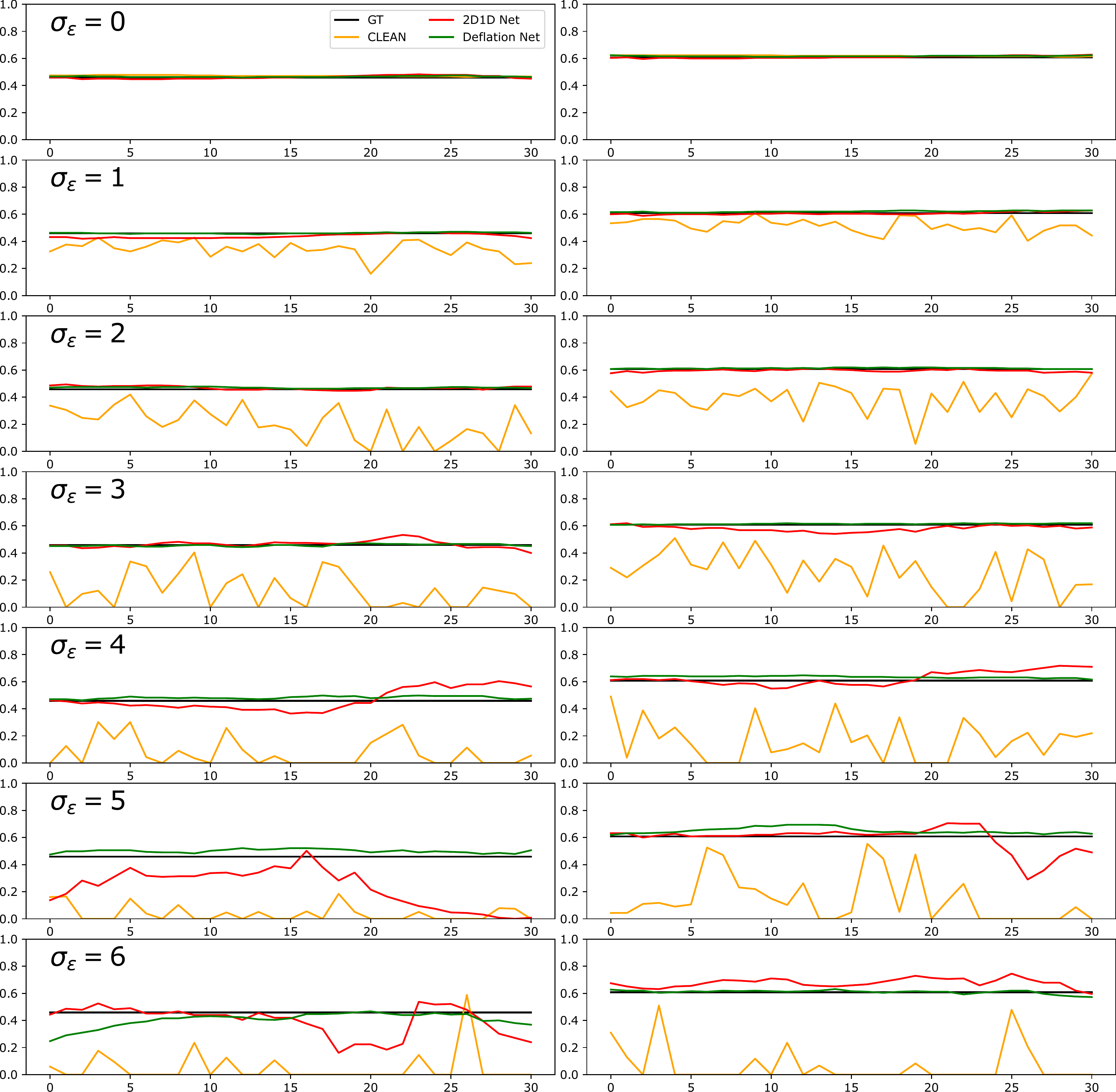}
\caption{Reconstructions of temporal profiles of constant sources. The horizontal and vertical axes indicate the time step and \bcf{normalized} amplitude for each subpanel.}
\label{transitcste}
\end{figure}

\begin{figure*}[!h]
\centering
\includegraphics[width=2\columnwidth]{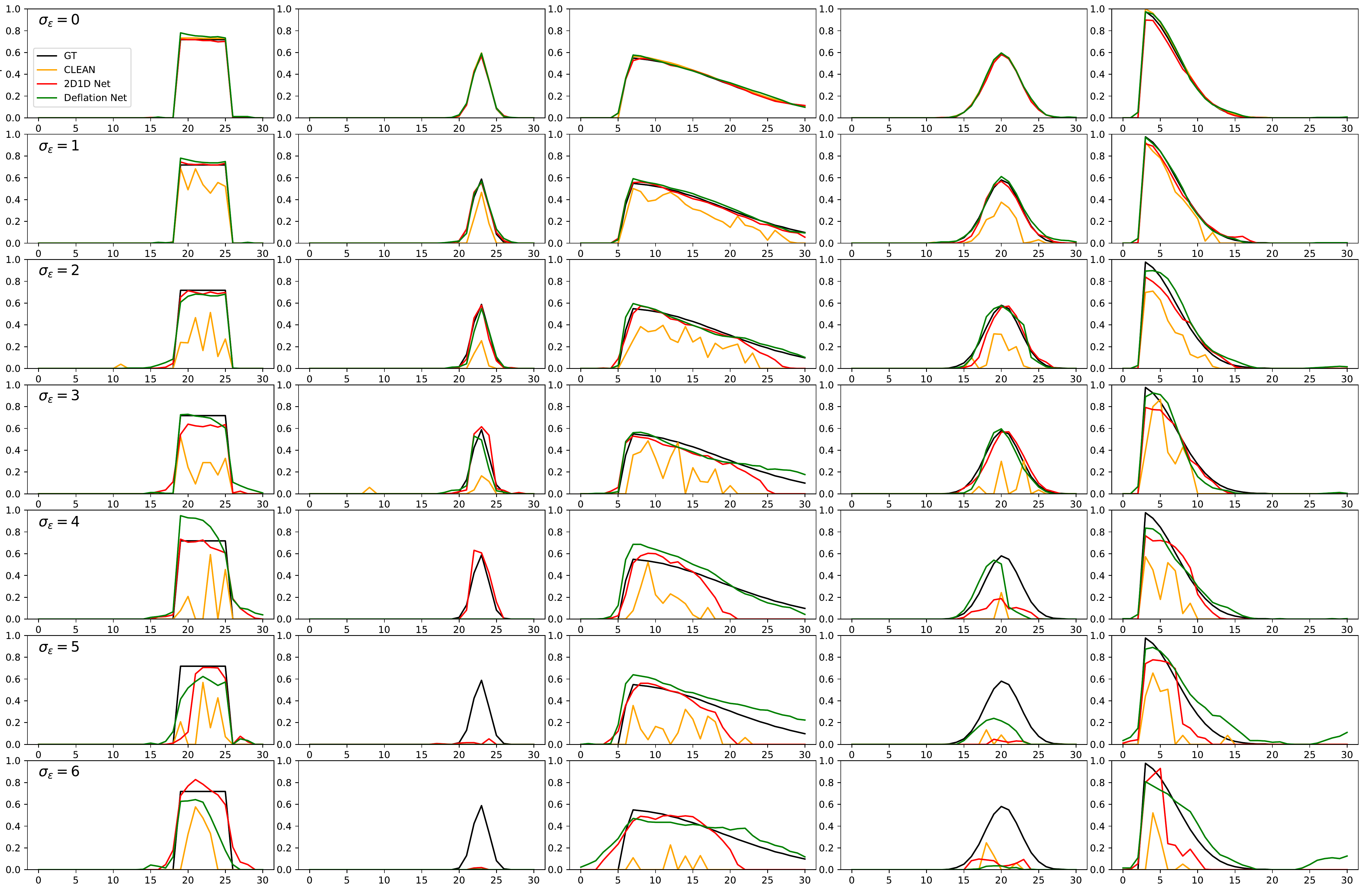}
\caption{Reconstructions of temporal profiles related to some transient sources in the test cubes. The horizontal and vertical axes indicate the time step and \bcf{normalized} amplitude for each subpanel. The figure compares methods at different noise levels $\sigma_\epsilon$. The names of the sources are, from left to right: {source 1}, {source 2}, {source 3}, {source 4}, and {source 5}.}
\label{1transit}
\end{figure*}

\begin{figure*}[!h]
\centering
\includegraphics[width=2\columnwidth]{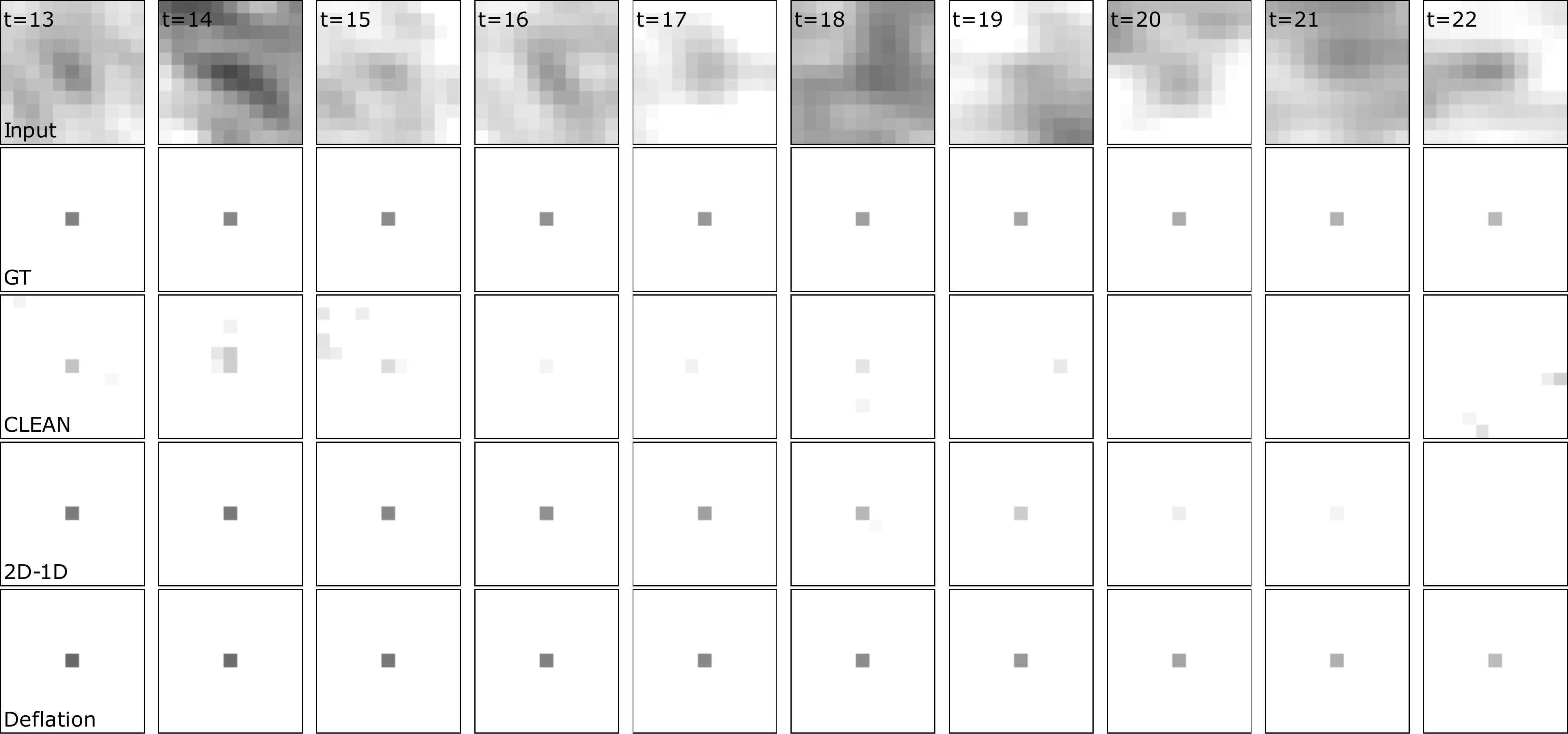}
\caption{Visualization of {source 3} in Figure~\ref{1transit} reconstructed from different methods, at noise level $ \sigma_\epsilon = 4 $.}
\label{18}
\end{figure*}

\begin{figure*}[!h]
\centering
\includegraphics[width=2\columnwidth]{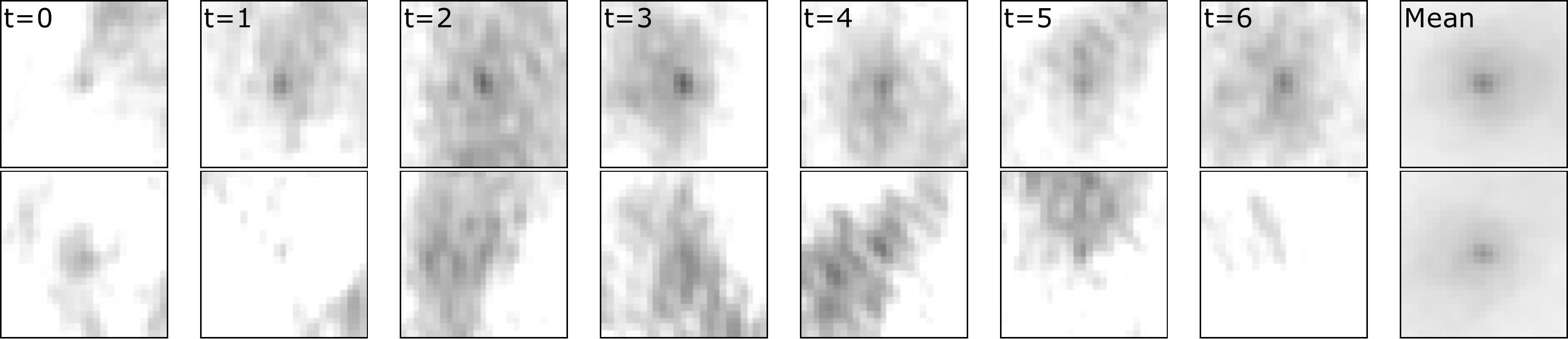}
\caption{Constant source in the input degraded sky between time steps 0 and 6, and average of the input skies over all the time steps. First row: noise level $ \sigma_\epsilon = 2 $. Second row: $ \sigma_\epsilon = 3 $. The noise level is reduced in the averaged sky.}
\label{averagesky}
\end{figure*}

\begin{figure}[!h]
\centering
\includegraphics[width=1\columnwidth]{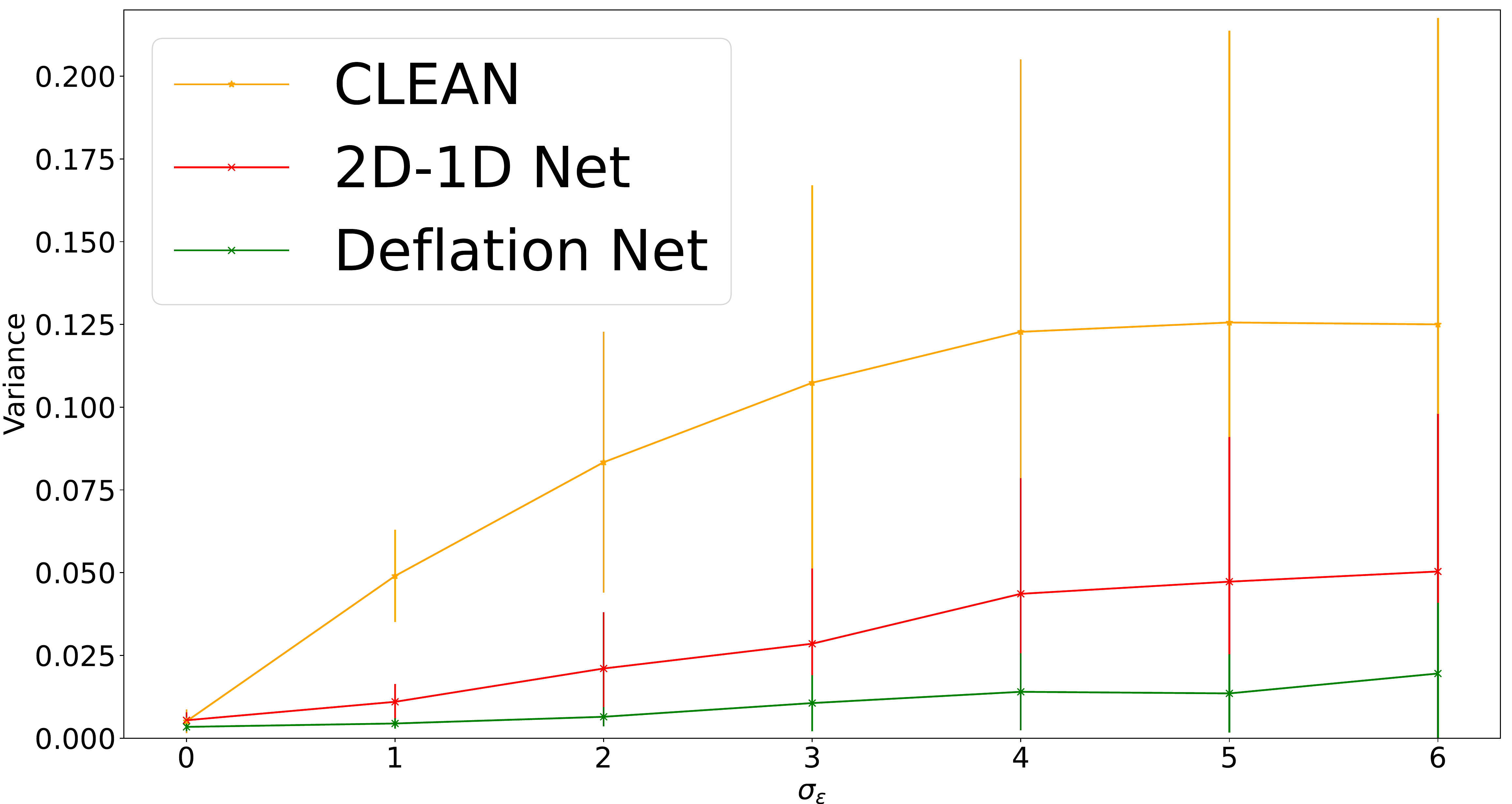}
\caption{Average variance of reconstructed light curves of constant sources over the test cubes at different values of the noise level $ \sigma_\epsilon $. The variances are computed on cubes normalized between 0 and 1. The vertical bars show the standard deviation.}
\label{variance}
\end{figure}

\begin{figure}[!h]
     \centering
     \begin{subfigure}[b]{1\columnwidth}
         \centering
\includegraphics[width=1\columnwidth]{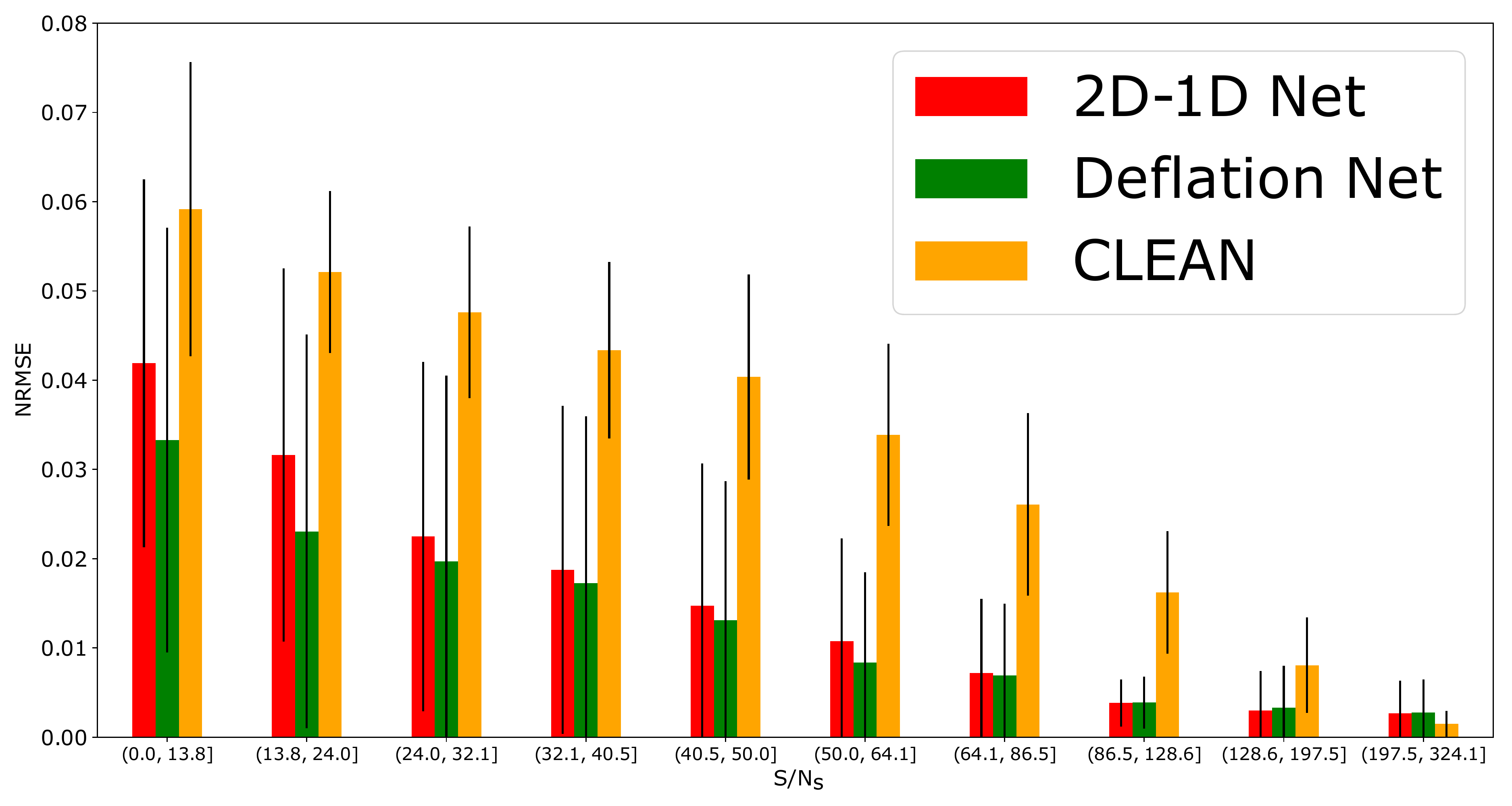}
     \end{subfigure}
     \begin{subfigure}[b]{1\columnwidth}
         \centering
\includegraphics[width=1\columnwidth]{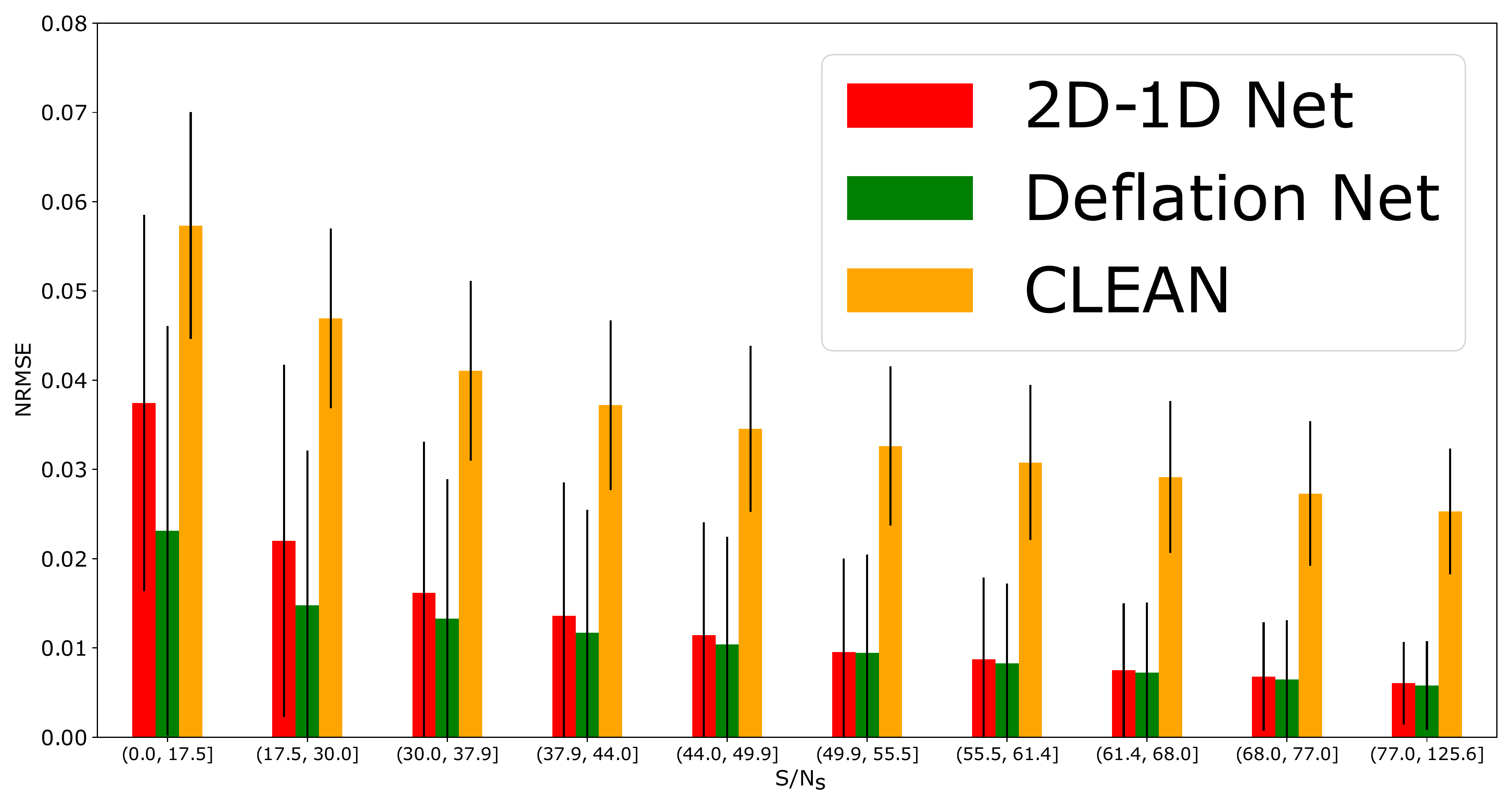}
     \end{subfigure}
        \caption{\bc{Average source reconstruction performance of 2D-1D Net, Deflation Net, and CLEAN, over the test sky cubes with (upper subpanel) a fixed test PSF cube and the noise levels $\sigma_\epsilon \in  \{0, 1, 2, 3, 4, 5, 6\} $; (lower subpanel) a fixed noise level $\sigma_\epsilon = 3 $ and the 50 test PSF cubes. Regarding all of these inferences, sources in the test sky cubes have different input $\mathrm{S/N_s}$. These $\mathrm{S/N_s}$ values can be divided into ten deciles. This figure visualizes $\mathrm{NRMSE_s}$ averaged over sources in each decile. The black whiskers indicate the standard deviations.}}
        \label{fig:three graphs 1}
\end{figure}

\begin{figure}[!h]
\centering
\includegraphics[width=1\columnwidth]{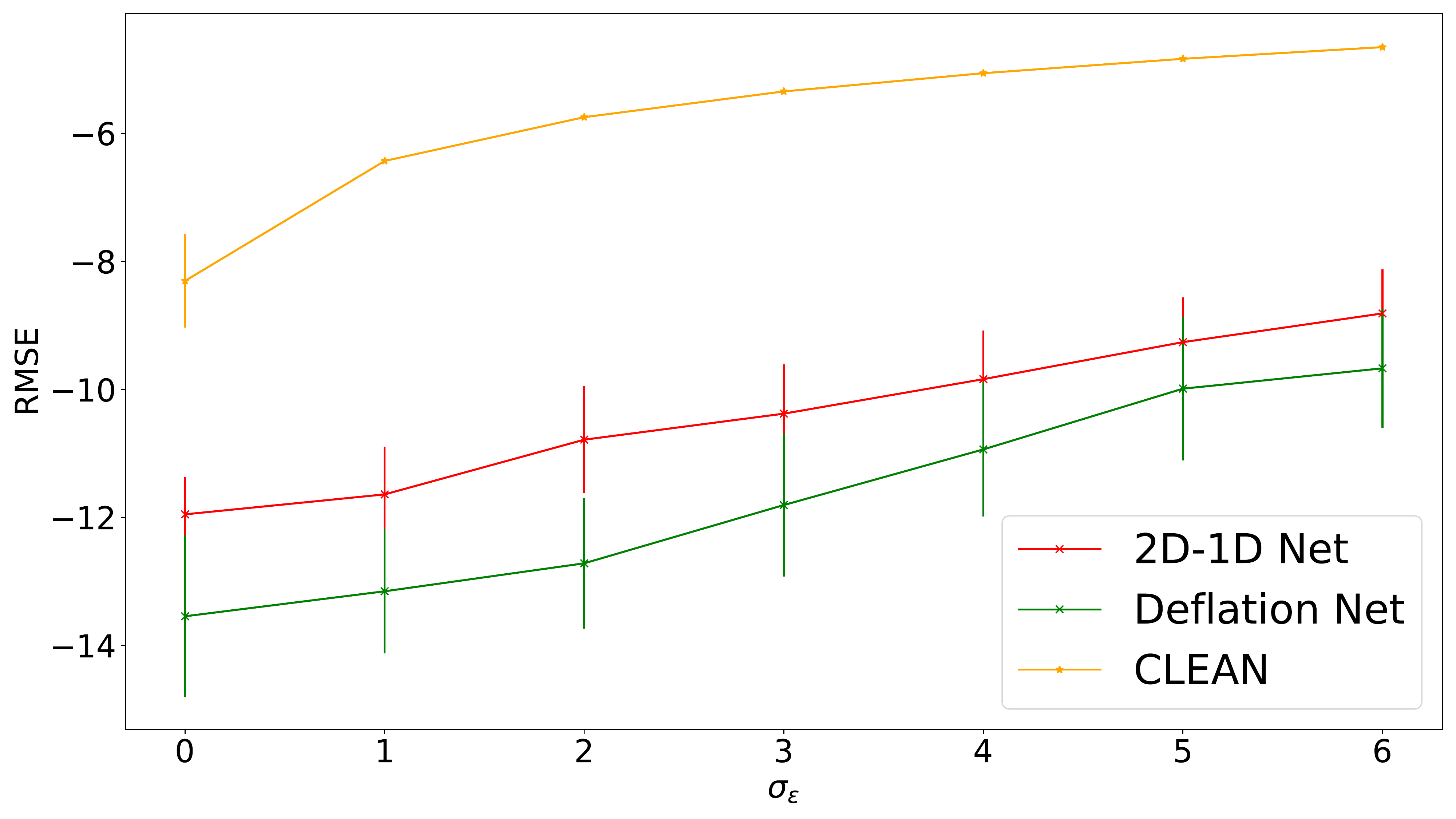}
\caption{$\log (\mathrm{RMSE}_\mathrm{noise})$ averaged over the test cubes at different values of $\sigma_\epsilon$. The vertical bars show the standard deviation. Because of the log scale, the standard deviation is higher when $\mathrm{RMSE}_\mathrm{noise}$ is small. This is the case for the DL-based methods. 
}
\label{backnoise}
\end{figure}

Figure~\ref{1transit} compares the reconstruction of temporal profiles related to several transient sources in the test cubes. The figure compares methods at different noise levels $\sigma_\epsilon$. We observe that DL-based methods produce better reconstructions than CLEAN. With increasing noise level $\sigma_\epsilon$, the performance of CLEAN rapidly degrades, whereas that of DL-based methods remains good. The plots of Figure~\ref{1transit} related to {source 1} illustrate this. Moreover, even in the presence of noise, DL-based methods provide a better restoration and preservation of the start and end dates of the transient signal. This is granted by the temporal modeling capability of these methods. Conversely, in the case of CLEAN, the restored start and end dates of the transient is distorted because of the noise in single frames. The plots related to {source 5} at noise level $ \sigma_\epsilon = 3 $ and $ \sigma_\epsilon = 4 $ illustrate these statements regarding the end of the transient signal. Among DL-based methods, Deflation Net seems better for restoring transient signals at a high-noise regime. The plots of {source 3} depict this. This suggests that the architectural choice of Deflation Net decoupling constant and transient source reconstructions is appropriate with respect to the inverse problem. Even at the highest noise level $ \sigma_\epsilon = 6 $, Deflation Net can restore the end of the transient signal. Figure~\ref{18} visualizes {source 3} reconstructed from different methods, at noise level $ \sigma_\epsilon = 4 $. We see that DL-based methods more efficiently reconstruct the transient source than CLEAN. Moreover, Deflation Net performs the best in reconstructing the end of the transient source. 
These results are particularly important whenever one telescope wants to react quickly upon detecting a potential transient. A transient could be detected on a  nearly real-time imaging pipeline that integrates our trained network as soon as its light curve rises. Alerts could therefore be created and distributed earlier and with more confidence concerning false detections.

As a matter of comparison and robustness, Figure~\ref{transitcste} compares reconstructions of temporal profiles for constant sources. As expected, DL-based methods systematically present reconstructions with higher fidelity. Moreover, they produce light curves with more stable amplitude variations than the CLEAN-based method. We must ensure that the trained network correctly reconstructs the light curves of constant sources without making them look like flickering transient sources. 
Moreover, Deflation Net presents less varying reconstructed constant sources than 2D-1D Net.  Deflation Net is understood to provide a better reconstruction of constant sources for the following reasons.  First, the noise level is reduced in the average input sky, and the average PSF is better conditioned than the individual PSF frame. Deconvolution of an average sky by the average PSF is thus easier for constant sources. Figures~\ref{averagesky} and~\ref{averagepsfimage} illustrate this by respectively displaying the average input sky for various noise levels and the average PSF. Second, the skip connections of Deflation Net (line~\ref{summ} of Algorithm~\ref{alg:algorithmDeflation}) allow reconstructions at different time steps to be distributed near the average deconvolved sky. Figure~\ref{variance} compares the average variance of reconstructed constant sources over the test cubes at different values of $ \sigma_\epsilon $. The figure confirms that DL-based methods reconstruct constant sources with better fidelity than CLEAN, and that Deflation Net produces the lowest variances concerning constant sources. 
On average, constant sources reconstructed by 2D-1D Net present a variance $\sim$3.0 times smaller than those reconstructed by CLEAN. Deflation Net's variance is $\sim$8.6 times smaller than CLEAN's.  



After aggregating results of all inferences with all of the noise levels $\sigma_\epsilon \in  \{0, 1, 2, 3, 4, 5, 6\} $, the upper subpanel of Figure~\ref{fig:three graphs 1} shows $\mathrm{NRMSE_s}$ averaged over sources belonging to the same S/N interval delimited by deciles in $\mathrm{S/N_s}$. We observe that DL-based methods generally perform better than CLEAN, except when the $\mathrm{S/N_s}$ is very high ($\mathrm{S/N_s} > 197.5$). This case indeed corresponds to  that of $\sigma_\epsilon = 0$, for which CLEAN is optimal. In other cases, DL-based methods, on average, perform better than CLEAN. By comparing positions of error bars indicating standard deviations, we observe that in many cases ($\mathrm{S/N_s}$ between 40.5 and 128.6), DL-based methods present inhomogeneous performance over different sources, but their worst performance is statistically still better than the best performance  of CLEAN. At a high-noise regime ($\mathrm{S/N_s}$ below 40.5), DL-based methods perform better than the CLEAN-based method on average. Within the DL-based methods, Deflation Net performs better than 2D-1D Net on average in most noisy cases (S/N between 0 and 86). On average, CLEAN presents $\mathrm{NRMSE_s}$ values that are $\sim$2.4 times higher than 2D-1D Net and $\sim$2.7 times higher than Deflation Net.

Figure~\ref{18} illustrates the high performance of DL-based methods in background denoising. We observe that they effectively restore the empty sky around the sources. This is less the case for the frame-by-frame CLEAN method, which  captures noise and generates residual noisy pixel distributions around the true source. Figure~\ref{backnoise} compares $\mathrm{RMSE}_\mathrm{noise}$, averaged over the test cubes at different values of $\sigma_\epsilon$. We observe that for all values of $\sigma_\epsilon$, DL-based methods better suppress background noise than CLEAN. Within the DL-based methods, Deflation Net better restores the background sky  because the noise level is reduced and the PSF is better conditioned in averaged input sky. On average, CLEAN presents $\mathrm{RMSE}_\mathrm{noise}$ values that are $\sim$1.8 times higher than the 2D-1D Net values and $\sim$2 times bigger than the Deflation Net values.







\begin{figure}[!h]
\centering
\includegraphics[width=1\columnwidth]{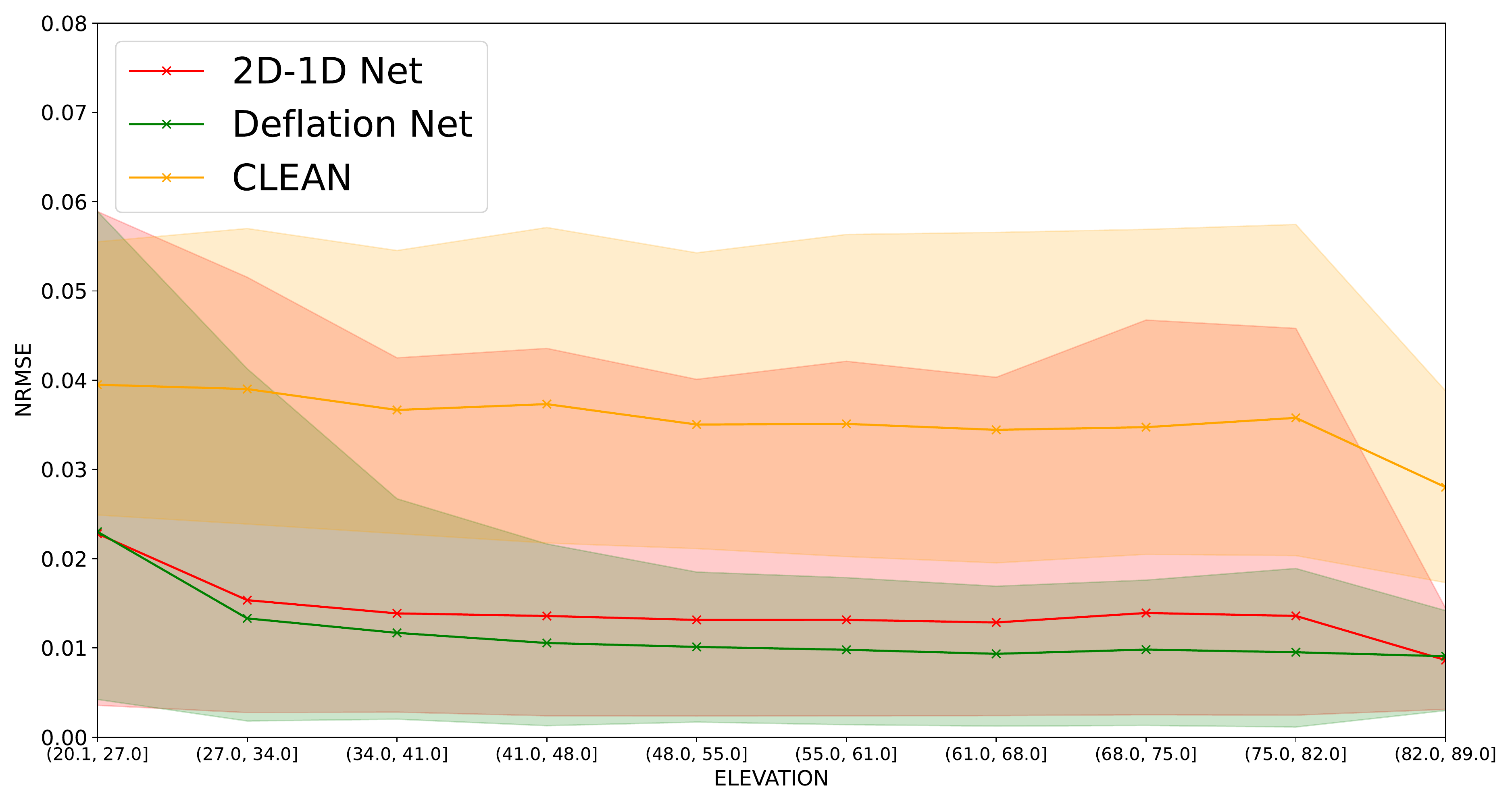}
\caption{Mean $\mathrm{NRMSE_s}$ for sources in the test set for ten equally spaced bins of PSF pointing elevations. The envelope for each method delimits the first and the last ten percentiles.}
\label{barelevenvelop}
\end{figure}



\subsection{Varying test PSF cube and fixed noise} \label{varypsfsec}

In this section we set $\sigma_\epsilon = 3$ and aggregate results over all of the PSF test cubes.


Figure~\ref{barelevenvelop} shows average $\mathrm{NRMSE_s}$ for sources in the test set for ten equally spaced bins of PSF elevations.  The apparent envelope for each method delimits the first and the last ten percentiles. We see that for all bins, our DL-based methods, on average, outperform CLEAN in large margins. This is especially true for Deflation Net, which presents the last ten percentiles near the first ten percentiles of CLEAN in most cases (elevations$>$41 degree). 
Among the  DL-based methods, here again, in most cases Deflation Net performs better on average than 2D-1D Net (elevations between 27 and 82 degrees). This confirms that  the Deflation Net architecture efficiently tackles the inverse problem. On average, CLEAN presents $\mathrm{NRMSE}$ values that are $\sim$2.6 times bigger than 2D-1D Net and $\sim$3.2 times bigger than Deflation Net. Furthermore, in most cases (elevations $>$ 34 degrees), the envelope of Deflation Net is smaller than those of 2D-1D Net and CLEAN. Therefore, Deflation Net brings less error dispersion in source light curve reconstruction than other methods. This can be explained by its skip connections that set reconstructions at different time steps near the average deconvolved sky. 
Our methods perform worse at elevations between 20 and 27 degrees and better at elevations between 82 and 89 degrees. Between these bounds, their performance seems to increase with increased elevation. This is coherent: the closer to 90 degrees the elevation, the better conditioned the PSF is, and therefore the easier it is to deconvolve. Conversely, lower elevation means that the PSF is worse conditioned and undergoes a strong projection effect, making the deconvolution task harder.

\begin{figure}[!h]
\centering
\includegraphics[width=1\columnwidth]{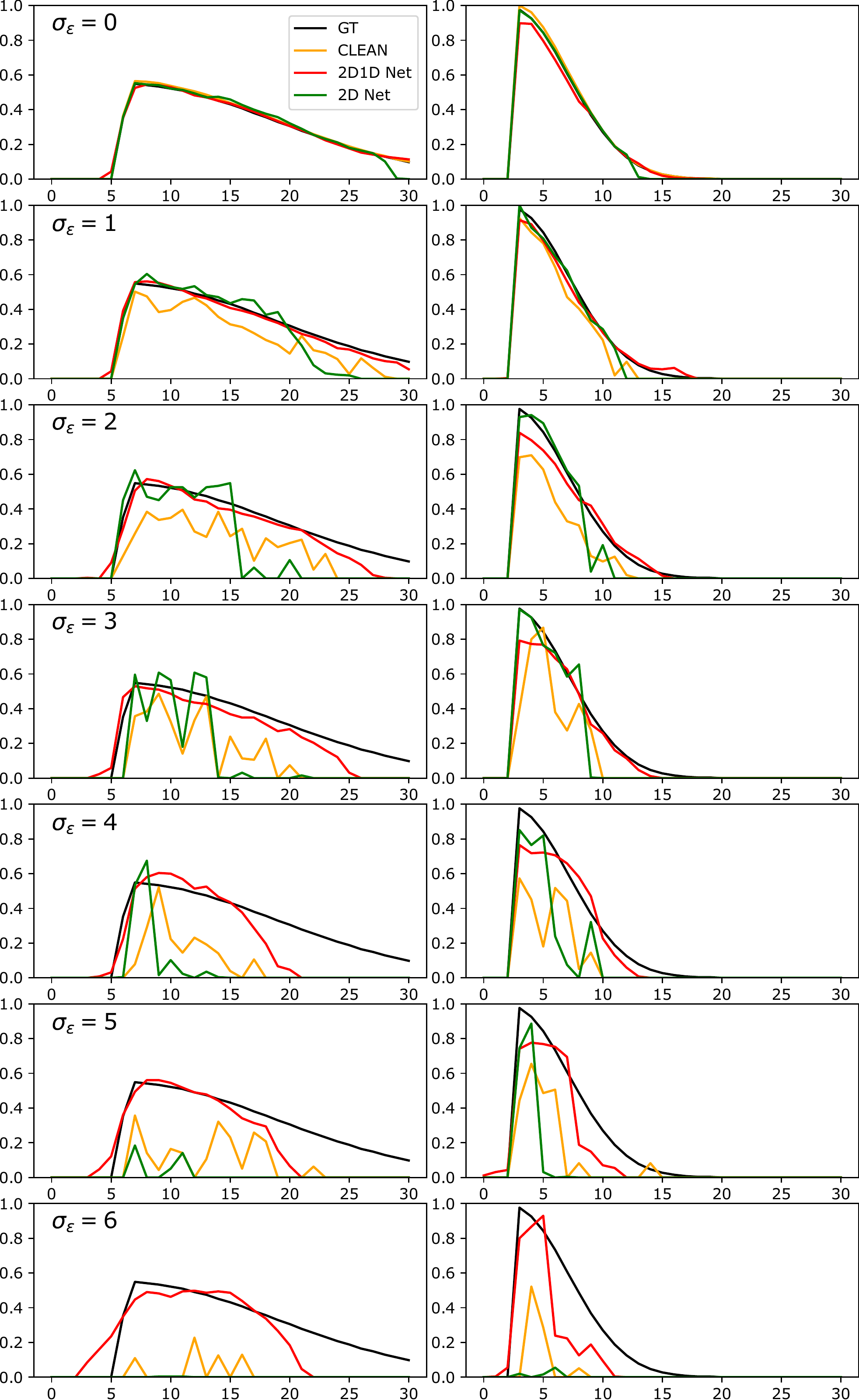}
\caption{Reconstructions of temporal profiles of {source 3} and {source 5} (also shown in Figure~\ref{1transit}) \bcf{at different noise levels $\sigma_\epsilon$}, from the \bcf{normalized} test cubes.} 
\label{1transitwabl}
\end{figure}

\begin{figure*}[!h]
\centering
\includegraphics[width=2\columnwidth]{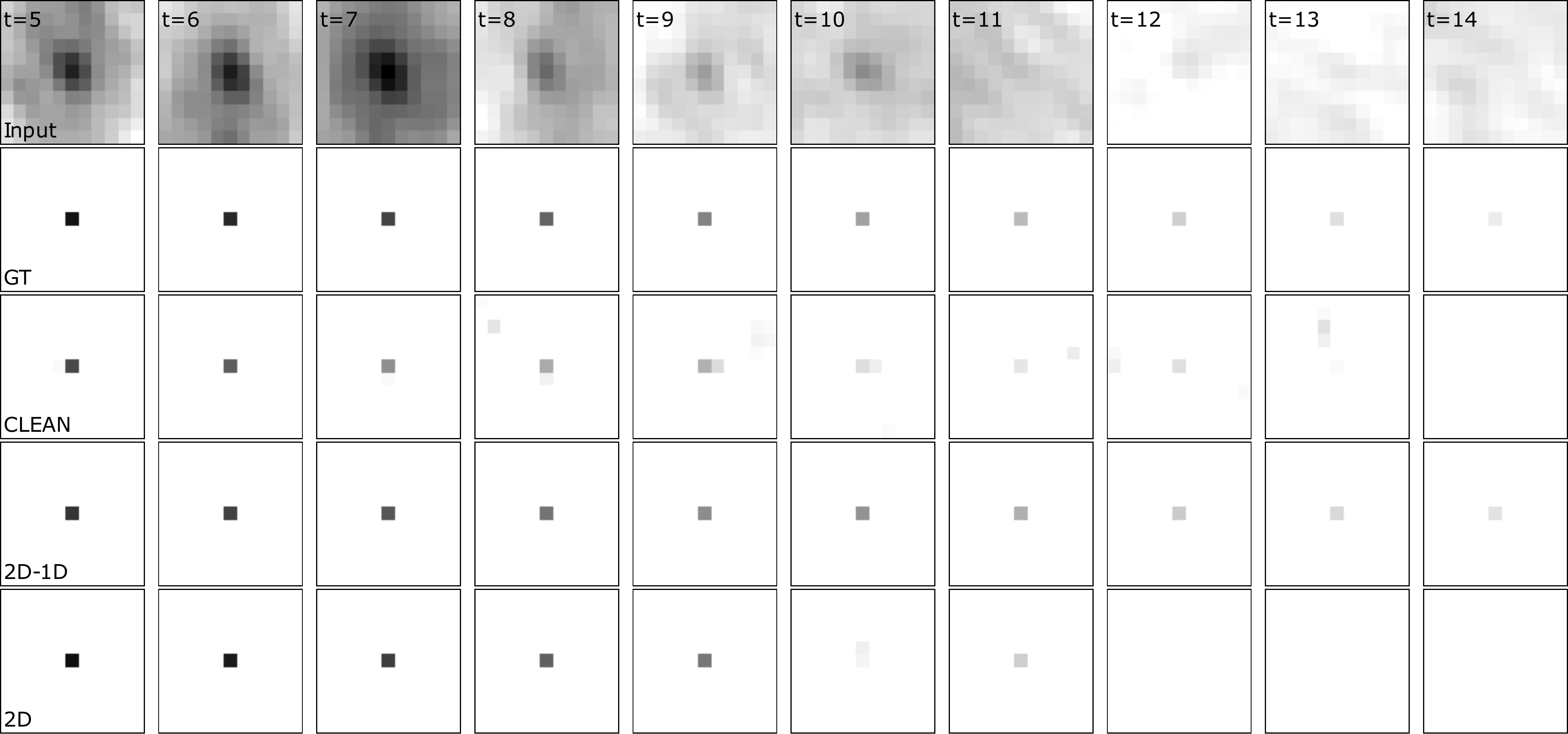}
\caption{Visualization of {source 5} in Figure~\ref{1transitwabl} reconstructed from different methods, at noise level $ \sigma_\epsilon = 2 $.}
\label{120}
\end{figure*}

Second, after aggregating the results of all inferences with all the test PSF cubes over the test sky cubes, the lower subpanel of Figure~\ref{fig:three graphs 1} shows $\mathrm{NRMSE_s}$ averaged over sources belonging to the same S/N interval delimited by deciles in $\mathrm{S/N_s}$. We observe that DL-based methods perform better on average than CLEAN for all intervals of $\mathrm{S/N_s}$. We can state the following by observing the black whiskers indicating standard deviations around mean values: in most cases, DL-based methods present inhomogeneous performance, but their worst performance is statistically still better than the best performance of CLEAN. Within the DL-based methods, Deflation Net performs slightly better than 2D-1D Net when $\mathrm{S/N_s}$ is higher than 44, and outperforms 2D-1D Net when $\mathrm{S/N_s}$ is below 44. This confirms again that the Deflation Net architecture is appropriate regarding the inverse problem. On average, 2D-1D Net presents $\mathrm{NRMSE_s}$ that is 3.10 times smaller than that of  CLEAN. Deflation Net shows $\mathrm{NRMSE_s}$ values that are 3.50 times smaller than that of CLEAN. 



\subsection{Importance of temporal modeling}

In this subsection we evaluate the benefit brought by the 1D Net in 2D-1D Net. This measures to what extent capturing the temporal structure of a signal increases the signal reconstruction performance. To do so, we compare the performance of CLEAN, 2D-1D Net, and 2D Net. 
The last is a variant of 2D-1D Net from which we excluded the 1D Net part, and we made the last 2D convolutional layer of 2D Net output a one-channel image at each time step. This mode reduces our method to a classic imager that forms a single image on time-integrated data.
Figure~\ref{1transitwabl} compares the reconstructions of temporal profiles of some transient sources in the test cubes with the fixed PSF used in Section~\ref{fixedpsf} and illustrated in Figure~\ref{testpsf510}. As expected, we observe that with increasing noise levels, 2D Net rapidly loses the capability to reconstruct the ends of the light curve, which lodge at the noise level. This drawback is shared with CLEAN. On the contrary, 2D-1D Net succeeds in restoring them properly. Figure~\ref{120} visualizes {source 5} reconstructed from different methods, at noise level $ \sigma_\epsilon = 3 $. We see that 2D-1D Net reconstructs the transient profile more efficiently down to the disappearance. 2D Net fails to reconstruct it, similarly to CLEAN. These figures illustrate that, despite the noise, 1D Net allows correctly reconstructing the beginning and the end of a transient event by enabling temporal modeling.



\begin{figure}[!h]
     \centering
     \begin{subfigure}[b]{1\columnwidth}
         \centering
\includegraphics[width=1\columnwidth]{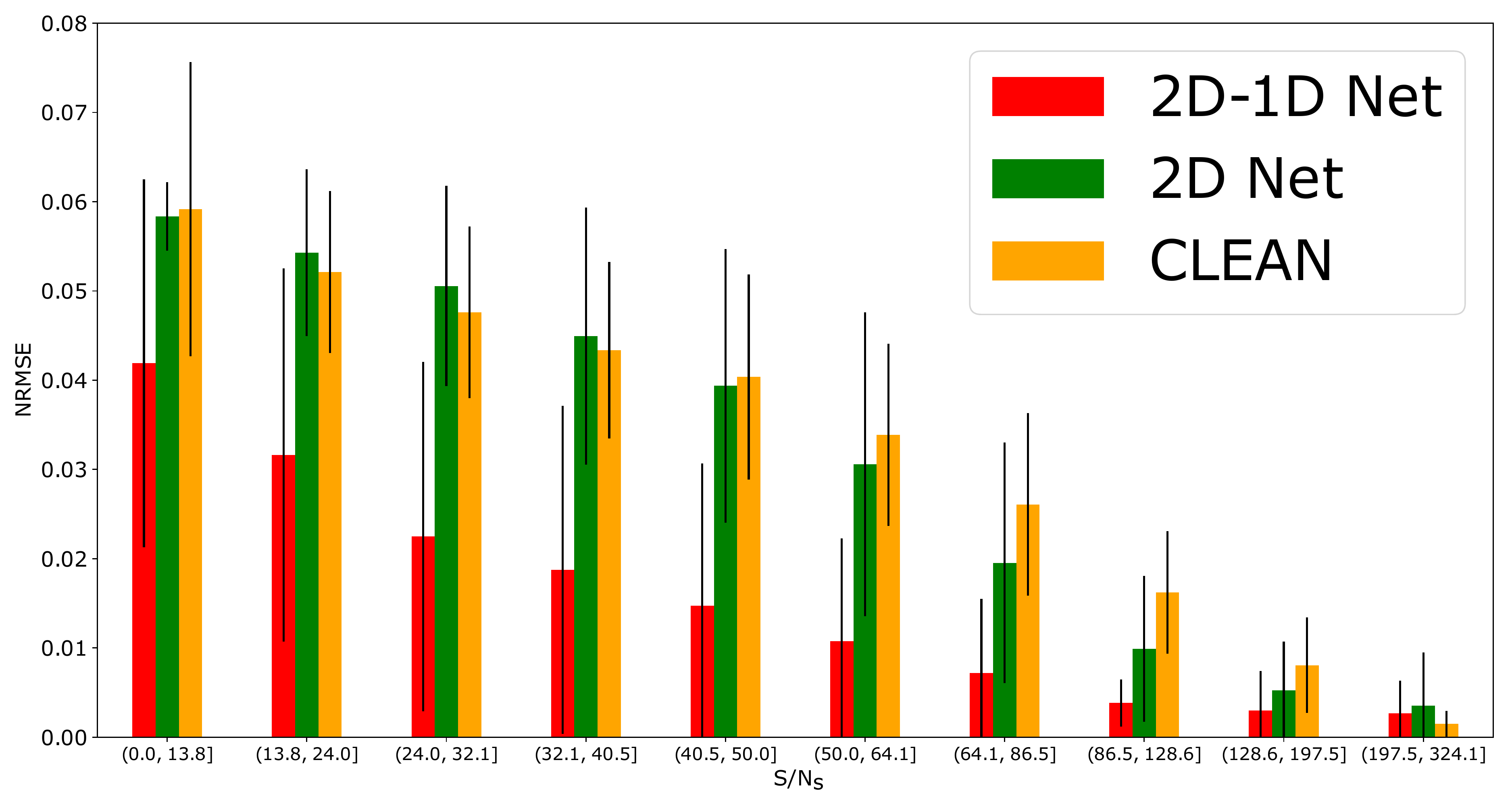}
     \end{subfigure}
     \begin{subfigure}[b]{1\columnwidth}
         \centering
\includegraphics[width=1\columnwidth]{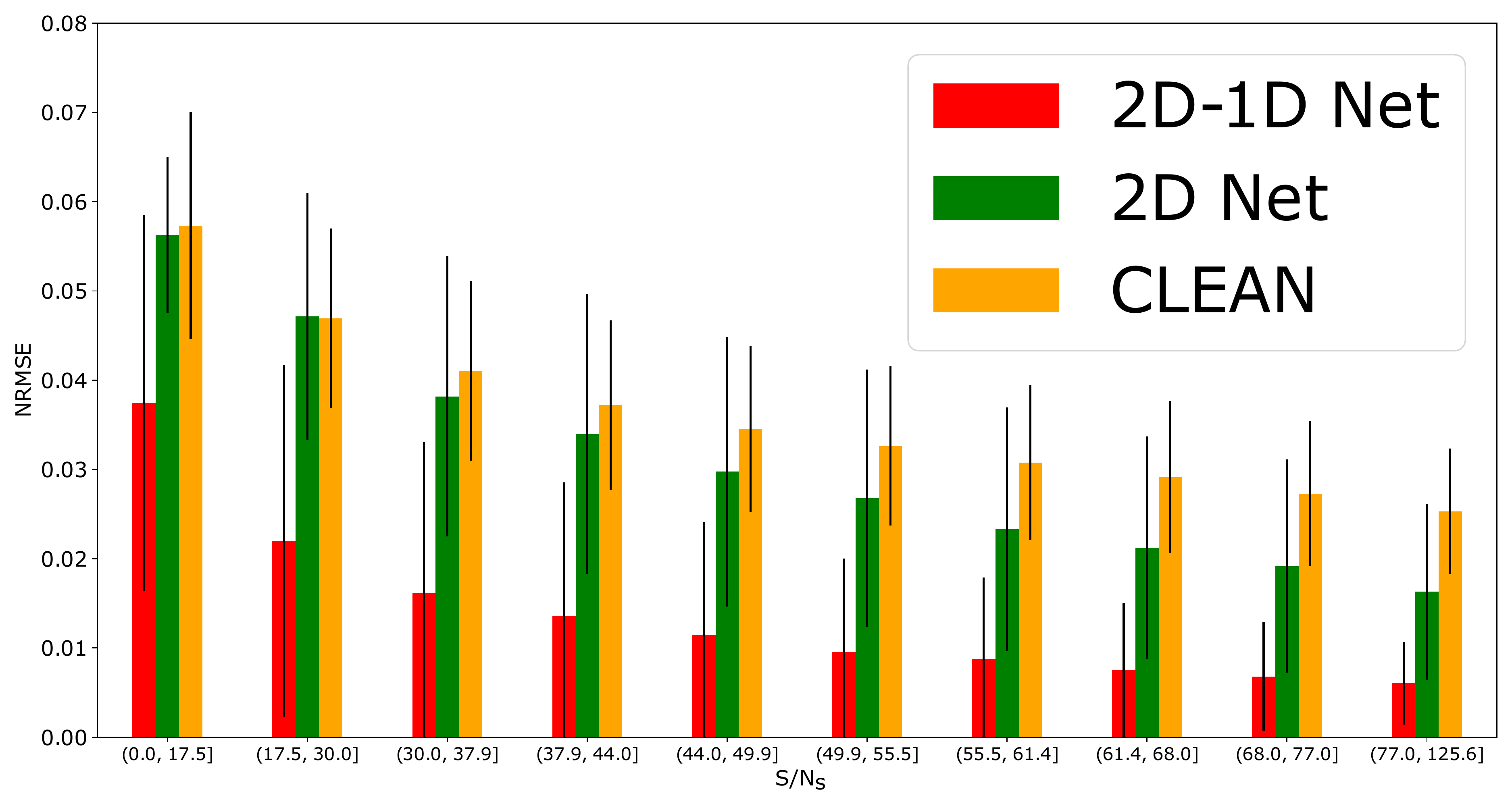}
     \end{subfigure}
        \caption{\bc{Average source reconstruction performance of 2D-1D Net, 2D Net, and CLEAN, demonstrating the importance of temporal modeling. These methods are evaluated on the test sky cubes with (upper subpanel) a fixed test PSF cube and the noise levels $\sigma_\epsilon \in  \{0, 1, 2, 3, 4, 5, 6\} $; (lower subpanel) a fixed noise level $\sigma_\epsilon = 3 $ and the 50 test PSF cubes. Regarding all of these inferences, sources in the test sky cubes have different input $\mathrm{S/N_s}$. These $\mathrm{S/N_s}$ values can be divided into ten deciles. This figure visualizes $\mathrm{NRMSE_s}$ averaged over sources in each decile. The black whiskers indicate standard deviations.}}
        \label{fig:three graphs 2}
\end{figure}

The upper subpanel of Figure~\ref{fig:three graphs 2} aggregates inference results on the test cubes with the fixed PSF cube illustrated in Figure~\ref{testpsf510} and the noise levels $\sigma_\epsilon \in  \{0, 1, 2, 3, 4, 5, 6\}$. The lower subpanel of Figure~\ref{fig:three graphs 2} aggregates inference results on the test cubes with the fixed noise level $\sigma_\epsilon = 3 $ and the 45 PSF test cubes used in Section~\ref{varypsfsec}. Both figures compare $\mathrm{NRMSE_s}$ averaged over sources of the same interval delimited by deciles in $\mathrm{S/N_s}$. On average, we observe that 2D Net performs slightly better than CLEAN in terms of $\mathrm{NRMSE_s}$. 2D-1D Net significantly outperforms both methods and presents on average $\sim 2.3$ times smaller $\mathrm{NRMSE_s}$ than 2D-1D Net and CLEAN. This exemplifies the benefit brought by the temporal modeling capability of 1D Net.



\begin{figure}[!h]
\centering
\includegraphics[width=1\columnwidth]{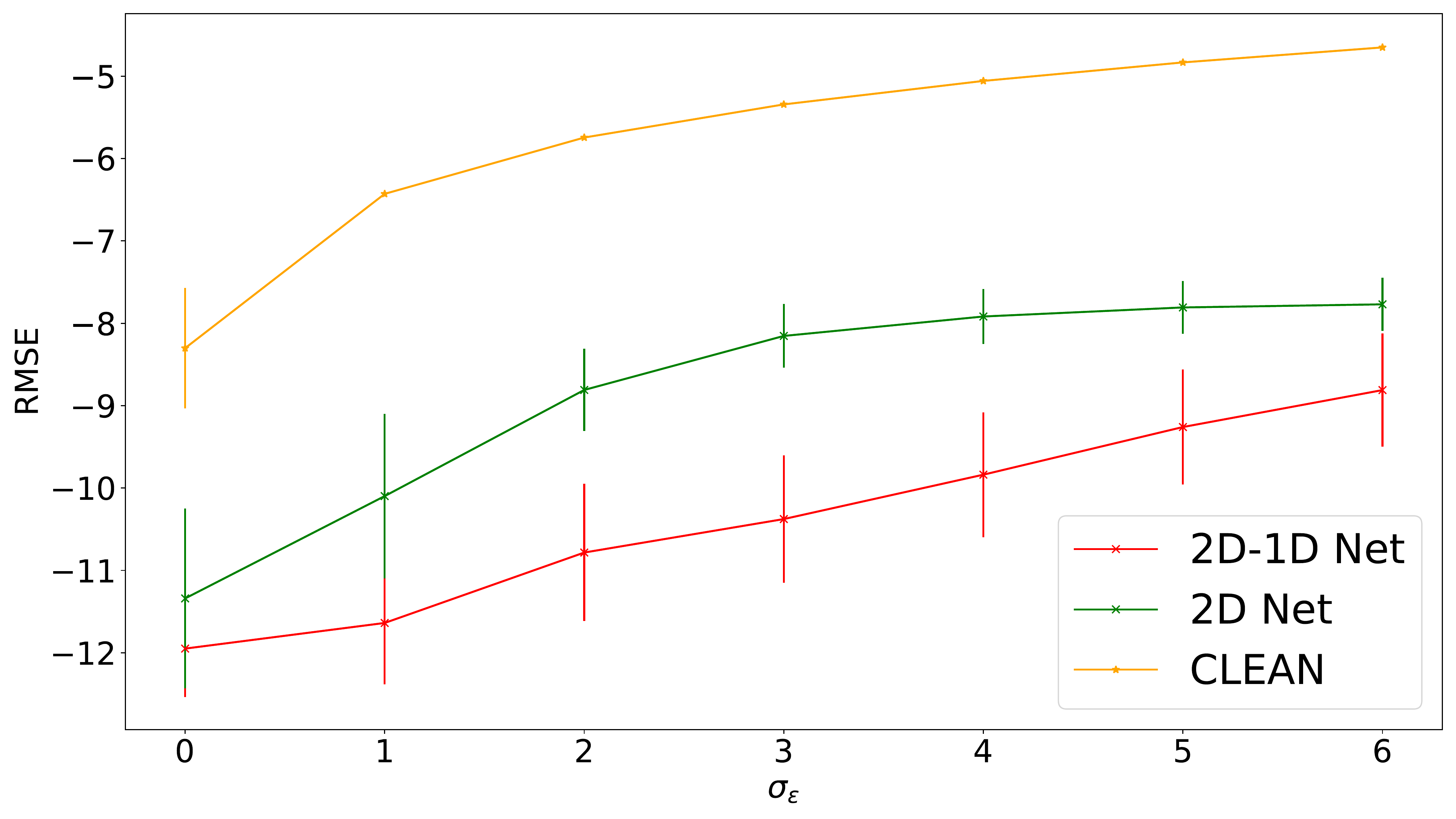}
\caption{$\log (\mathrm{RMSE}_\mathrm{noise})$ averaged over the test cubes at different values of $\sigma_\epsilon$. The vertical bars show standard deviation. Because of the log scale, the standard deviation is higher when $\mathrm{RMSE}_\mathrm{noise}$ is small. This is the case for the DL-based methods.}
\label{backnoisewabl}
\end{figure}

Even if 2D Net performs similarly to CLEAN based on $\mathrm{NRMSE_s}$, the former performs better regarding background denoising. This is shown in Figure~\ref{backnoisewabl}, which compares $\mathrm{RMSE}_\mathrm{noise}$, averaged over the test cubes at different values of $\sigma_\epsilon$ and the PSF illustrated in Figure~\ref{testpsf510}. For all values of $\sigma_\epsilon$, 2D Net better suppresses background noise than CLEAN by large margins. This shows that even if 2D Net does not extract temporal features, its convolutional layers enable highly efficient denoising. 2D-1D Net performs even better than 2D Net regarding this background denoising task. This shows that the temporal modeling capability of 1D Net also contributes to increasing the background denoising performance. On average, 2D Net presents $\mathrm{RMSE}_\mathrm{noise}$ that are $\sim 1.5$ times smaller than that of  CLEAN. 2D-1D Net presents $\mathrm{RMSE}_\mathrm{noise}$ that are $\sim 1.2$ times smaller than 2D Net. Figure~\ref{120} illustrates these observations. Especially at the end of the transient event, frame-by-frame CLEAN captures noise and generates residual noisy pixels around the true source. 2D Net generates less noisy pixels (it only generates a noisy pixel at time step $t = 10$). 2D-1D Net does not generate any obvious noisy pixel.

\section{Discussion and limitations}
\label{Sect:Discussion}

In this section, we discuss some relevant points, regarding our methods. The first point concerns the {support size of the frame}. In our inverse problem, we approximate the aperture synthesis method as  linear degradation embodied by a multiplication of the true data with a mask in the Fourier domain. Moreover, we carried out our study in the gridded space where all quantities are discrete set with constant frame support size. The mask, therefore, has the same support size as the sky. All frame support sizes between the sky, the mask, and the PSF cubes are linked. Therefore, if one wants to change the support size, new training of the networks is required. This is contrary to some other DL-based image or video restoration problems where the linear degradation is defined by a convolution rather than by a mask. In this case, a trained neural network can deconvolve an image of any size. Regarding our networks, as only the 2D Net module operates deconvolution, we should only retrain this module, and 1D Net can be frozen. From a dataset of ungridded visibilities, it is up to the scientist to decide which support size fits the scientific objective of the data. It should be trained on a support size that corresponds to the final products used (e.g., image catalog of a survey, transient detection pipeline).

Secondly, the {total duration of transient events} is worth considering. The total observation period and the number of temporal frames (which defined a slice PSF of the PSF cube) determine the upper limits on the entire duration and the shortest timescale of transient events we deal with. As an illustration, for a fixed number of PSF frames over the total observation period, an increase in the latter will increase the integration time of each PSF frame. Thus, if the total observation period is more extended, with a fixed number of PSFs, each PSF   appears smoother and presents fewer secondary lobes. Therefore, each time the total observation period changes, our networks should be theoretically retrained with new PSF cubes. We can suppose that the profiles of transient events in the train sky cube are diverse enough to be representative of real transients that can be encountered. In this case, we can keep the same train sky set. Conversely, if we suppose the length of the observation increases for a fixed PSF time integration, the third dimension of the cube increases, but each PSF will share the same characteristics as before. In that case, we should only retrain 1D Net, and 2D Net can be frozen as its role is to deconvolve PSFs as was done previously.  

{Computational complexity} is another discussion point. Once trained, our networks present faster reconstruction than frame-by-frame CLEAN  because, contrary to CLEAN, they are not iterative algorithms and can benefit from efficient GPU-based architectures. This aspect makes them attractive to the radio astronomy community. On our Intel I9-10940X CPU and NVIDIA TITAN RTX GPU, 2D-1D Net reconstructs the sky cube of 32 frames in $\sim 65$ ms. This speed is $\sim 75$ ms for Deflation Net. Both of them contain 0.1 M parameters.

Regarding {theoretical guarantee}, one of the drawbacks of our DL-based methods is that they lack mathematical frameworks allowing us to derive the 
    theoretical error bounds or uncertainty beyond empirical statistics on the results. 

It is also possible to extend our proposed methods to the polychromatic case, where the inverse problem also has a frequency dependency. In this case, the dirty, GT, and PSF cubes are four-dimensional data structures, adding the frequency dimension. A simple way to realize this extension is   i) to stack several skies at different frequencies in the channel dimension at the input of the 2D Net module and ii) to stack several PSFs at different frequencies in the channel dimension at the input of the SFT layer. The computation of the average sky and PSF, and the subtraction step in the Deflation Net should be done for each frequency.

Our neural network systems considered normalized pixel values; the ground truth images’ pixel values were divided by 100 (i.e., they lie between 0 and 1). In the same spirit, when facing different pixel intensity ranges occurring in other types of datasets, for example the real one, a min-max normalization allows us to bring our methods into conditions similar to those that are tackled in this paper. The neural networks can eventually be retrained.

Finally, \bc{our proposed methods can  easily be plugged into modern imagers, such as those based on wide-field imaging using projection (e.g., WSCLEAN; \citealt{offringa2014wsclean}) or those based on faceting (e.g., DDFacet; \citealt{Tasse2018}). These imagers address direction-dependent effects corrections (i.e., the correction of the widefield projection effects or the non-coplanar arrays) and its counterpart (killMS) takes care of the calibration regarding these effects. While this is not in the scope of our paper, making a calibration tool based on CNNs is also an interesting topic to consider. It would require a full recasting of the problem in the framework of the Radio Interferometer Measurement Equation (RIME) \citep{smirnov2011revisiting}, including the relevant Jones matrices and using ungridded visibilities directly. We leave these considerations for future work.}

\section{Conclusion}
\label{Sect:Conclusion}
In this study we dealt with transient source reconstruction in the context of radio interferometric imaging. We formulated this problem as a deconvolution of image time series. We proposed two neural networks to address this task, namely 2D-1D Net and Deflation Net. Thanks to the SFT layer, they can handle multiple PSFs that vary depending on the observed sky positions. They involve the same submodules: 2D Net and 1D Net. The former deconvolves individual frames and the latter enables temporal sky modeling. 2D-1D Net is based on a simple feedforward inference, whereas Deflation Net involves different computational flows. The latter restores the average sky and uses it to isolate transient sources in individual frames. Experiments based on simulated data and metrics measuring temporal profile reconstruction and background denoising demonstrated superior performance of these DL-based methods over CLEAN in the presence of noise. 
Deflation Net performs the best, excelling in reconstructing constant sources and background denoising. The ablation study confirms the temporal modeling enabled by 1D Net significantly increases the sky cube reconstruction performance.

Learning methods adapted to deconvolution, temporal structures of transient sources, and specificities of instrumental response are key elements to efficiently analyzing the images obtained via radio interferometers in the SKA era. For instance, the raw sensitivity of MeerKAT enables deep imaging (i.e., a typical $\sigma\sim 10$ $\mu$Jy in 15 min) as well as high cadence imaging capabilities in a large field of view. Being able to deconvolve the data in high-noise regimes efficiently will maximize the chance to discover new transients and overcome the limitations imposed by current deconvolution methods. Missing transients in the image plane can be due to the lack of sensitivity (i.e., detection problem) or the lack of sufficient temporal sampling (i.e., dilution problem) that averages out short-scale transients. The robust 2D and 1D image reconstruction brought by the trained networks introduced in this study has significantly improved the 3D estimation of the sky. Replacing classical frame-by-frame deconvolution methods with the DL-based reconstruction methods can lead to better use of telescope time. This will drastically reduce the required exposure time while enabling a faster temporal sky sampling. 





\begin{acknowledgements}
      This work is supported by the European Community through the grant ARGOS (contract no. 101094354) and by Safran Electronics \& Defense.
\end{acknowledgements}

\bibliographystyle{aa}
\bibliography{references}

\begin{appendix} 

\section{Coordinates of an astrophysical source}

\begin{figure}[!h]
\centering
  \includegraphics[width=1\linewidth]{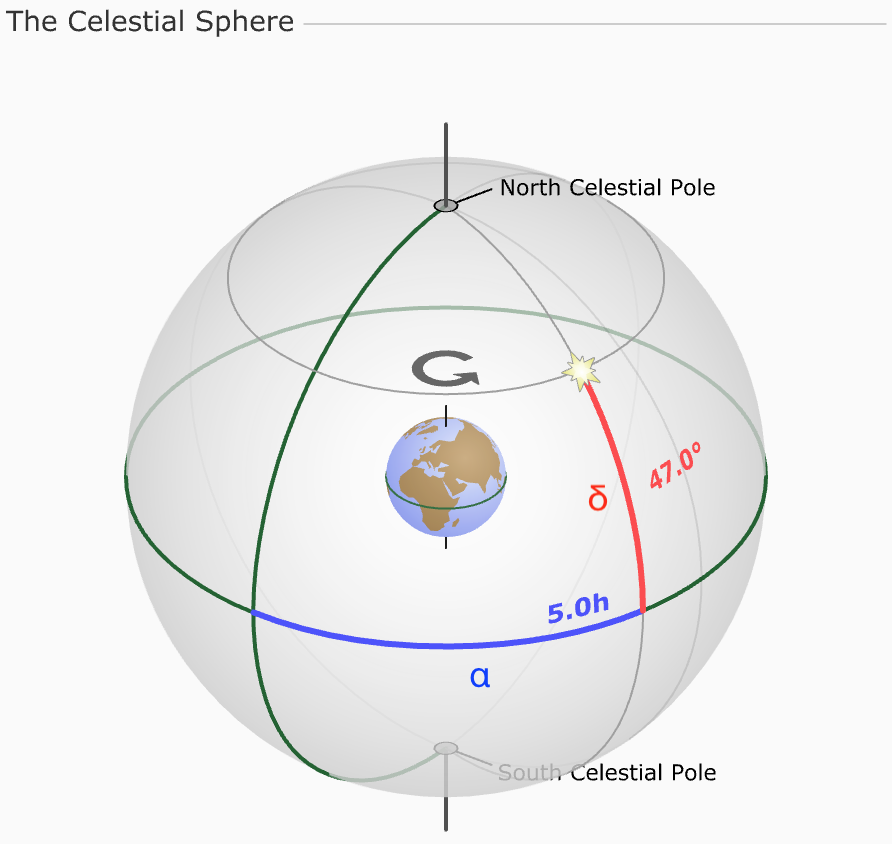}
\caption{\bc{Astrophysical source associated with a unique pair of coordinates (declination $\delta$ and right ascension $\alpha$) on the celestial sphere. Figure adapted from \citet{WinNT}.}}
\label{fig:coordinates}
\end{figure}

\section{Average PSF over the total observation duration}

\begin{figure}[!h]
\centering
\includegraphics[width=0.5\columnwidth]{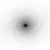}
\caption{\bc{Average PSF over the PSF cube in Figure~\ref{testpsf510}}. 
}
\label{averagepsfimage}
\end{figure}




% ...


\end{appendix}

\end{document}